\documentclass[prc,showpacs,floatfix,twocolumn,superscriptaddress,nofootinbib]
{revtex4}
\usepackage{colordvi}

\usepackage{graphics}
\usepackage{longtable}
\usepackage{epsfig}
\usepackage{amsmath,amssymb,amsfonts}

\begin{document}

\title{Transverse wobbling-a collective mode in odd-A triaxial  
nuclei }
\vskip 10pt

\author{S. Frauendorf}
\email{sfrauend@nd.edu} 
\affiliation{Department of Physics, University of Notre Dame, South Bend, Indiana 46556, USA}
\author{F. D\"onau}
\email{doenau@hzdr.de}
\affiliation{Institut f\"ur Strahlenphysik, Helmholtz-Zentrum Dresden-Rossendorf, 01314 Dresden, Germany}

\begin{abstract}
{The wobbling motion  of a triaxial rotor coupled to a high-j quasiparticle is treated semiclassically. Longitudinal and transverse coupling regimes  can be distinguished depending on, respectively, 
whether  the quasiparticle angular momentum is oriented parallel or perpendicular 
 to the rotor axis with the largest moment of inertia. Simple analytical  expressions for the wobbling frequency and the electromagnetic E2 and M1 transition probabilites are derived
assuming rigid alignment of the quasiparticle with one of the rotor axes and harmonic oscillations (HFA). Transverse wobbling is characterized  by a decrease of the wobbling frequency with increasing angular momentum.
Two  examples for transverse wobbling,  $^{163}$Lu and $^{135}$Pr,  are studied  in the framework of the full triaxial particle-rotor model 
and the HFA. The signature of transverse wobbling, decreasing wobbling frequency and enhanced E2 inter-band transitions, is found in agreement with experiment.  
}

\end{abstract}

\maketitle
\section{INTRODUCTION}

\begin{figure}
\includegraphics[height=6.5cm]  {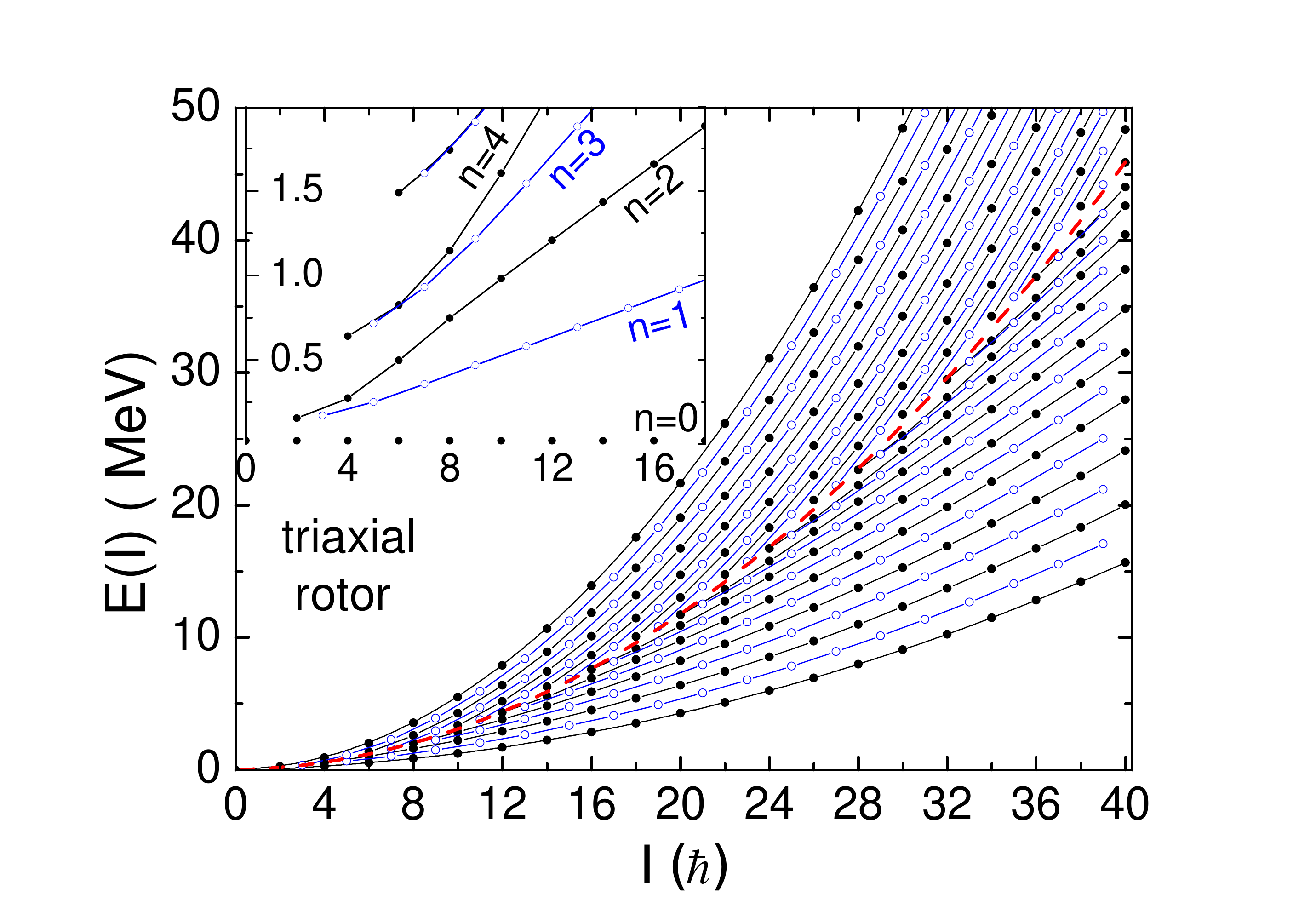} 
 \caption{\label{simpletriax} (Color online) 
Rotational band structures calculated for the triaxial rotor Hamiltonian, Eq.\,(\ref{rham}).
The inset shows a blow up of the energies of the lowest excited bands ($n=1-4$\,) relative to the yrast line ($n=0$). 
Full black dots belong to the states of signature $\alpha$=0 
and empty blue dots to signature $\alpha$=1.  The red line displays the separatrix. 
The ratios of the rotational  parameters  are $A_1=6A_3$ and
$A_2=3A_3$. The  energies are scaled such that $E(2_1^{+}$)\,=\,0.1 MeV.
}
\end{figure}

\begin{figure}
\hspace*{-1cm}
\includegraphics[scale=0.35]  {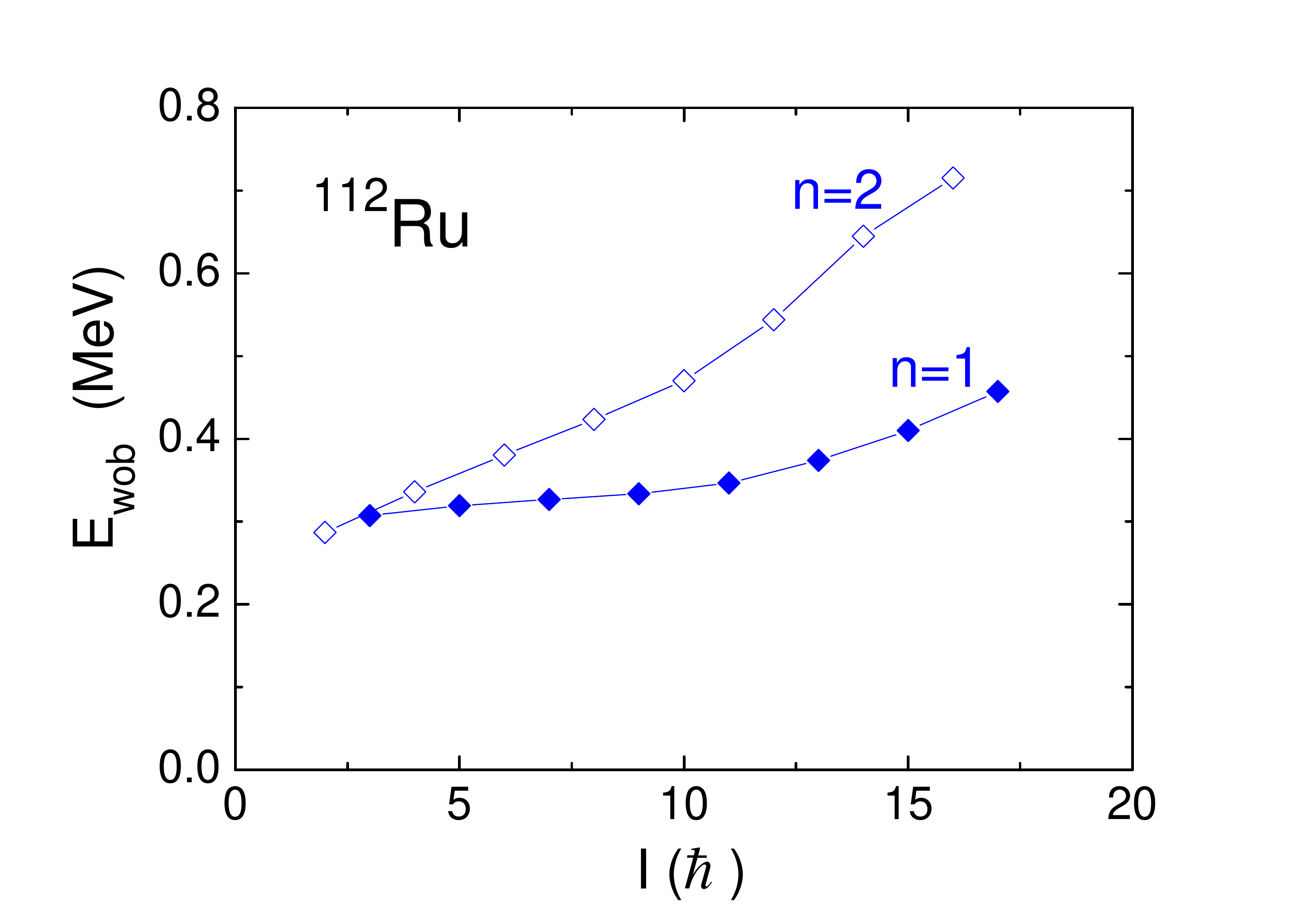} 
\caption{\label{ru112} (Color online)
Experimental energies  of the two lowest  wobbling bands $n = 1$, 2  relative to the  $n = 0$ yrast 
sequence in $^{112}$Ru. Data from \cite{Zhu09}}
\end{figure}

Textbooks on Classical Mechanics (see e.g. \cite{Landau}) discuss the motion of a rigid rotor with three different moments of inertia (MoI). Uniform 
rotation about the axis with the largest MoI corresponds to the lowest energy for given angular momentum (a.m.). For slightly larger energy, this axis executes
harmonic oscillations about the space-fixed a.m. vector. The frequency of these oscillations is proportional to the rotational frequency of the rotor. 
For quantal systems these oscillations appear as equidistant excitations. They were first observed in molecular spectra, and theoretically analyzed in Ref.\,\cite{MolWob}.    
Bohr and Mottelson applied  the concept to triaxial nuclei and introduced the name "wobbling" for the excitations. Fig.\,\ref{simpletriax} shows a triaxial rotor spectrum. 
With increasing spin $I$, the lowest excitations above the yrast states  become  more and more equidistant. This is the classical wobbling regime, which is characterized 
 by an increasing wobbling frequency with $I$ 
(we identify the experimental wobbling frequency $\hbar \omega_w$ with the energy difference   
$E(I)_{n=1}-\left(E(I+1)_{n=0}+E(I-1)_{ n=0}\right)/2$\,).
Such wobbling spectra have rarely been observed in nuclei.   The reason is that stable triaxial ground states are very
uncommon. Fig.\,\ref{ru112} shows the best example identified so far. 

\begin{figure}
\includegraphics[scale=0.35] {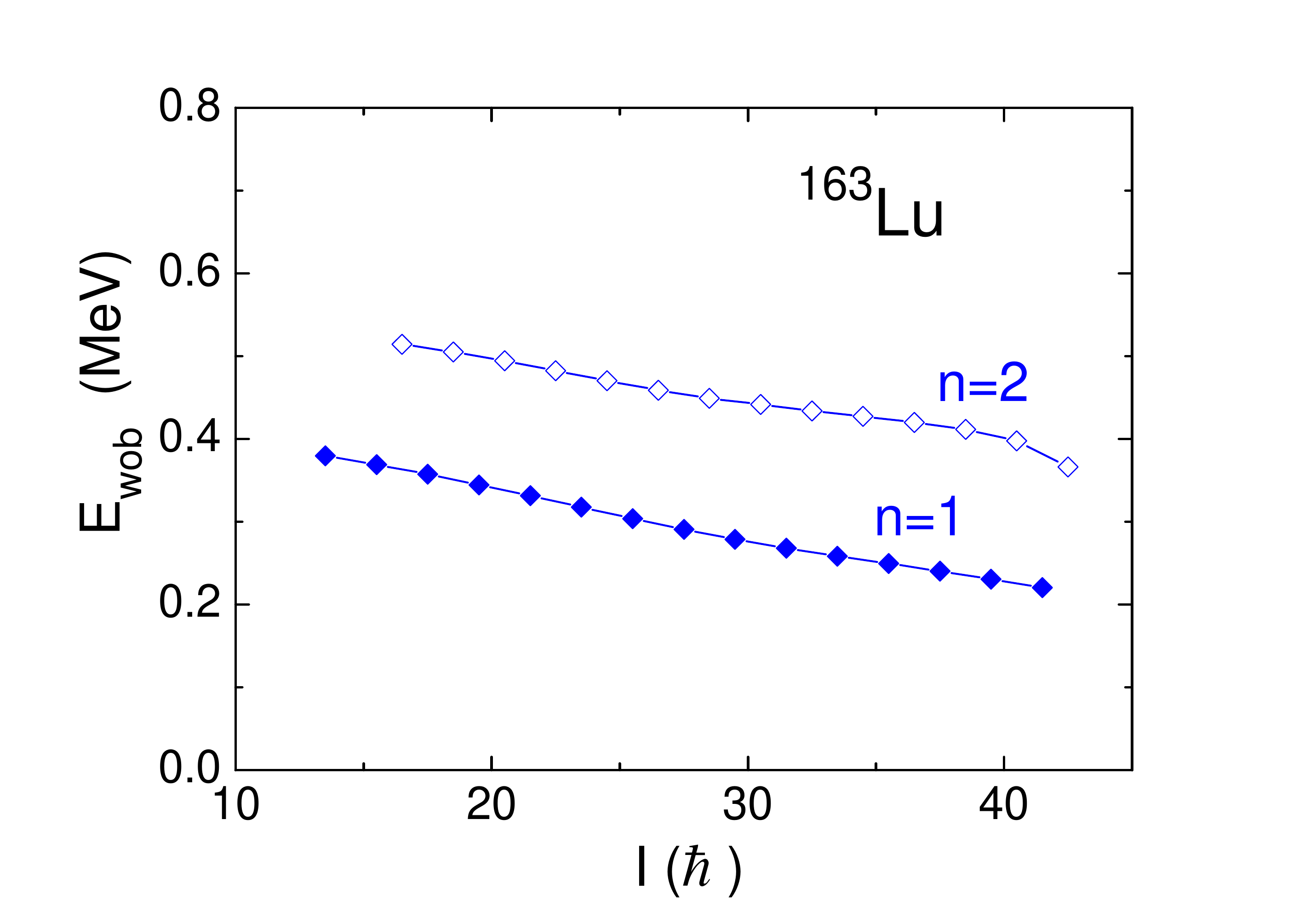} 
\caption{\label{lu163} (Color online) 
Experimental energies  of the two lowest wobbling bands $n = 1$, 2  relative to the $\pi i_{13/2}$ $n = 0$ 
sequence in $^{163}$Lu. Data from \cite{Od01}.
}
\end{figure}

The wobbling mode has been extensively studied for nuclei in the Triaxial Strongly Deformed (TSD) region 
 around $N=94$, where 
a significant gap opens in the neutron levels   at high spins for  
TSD shapes with $\varepsilon\approx 0.4$ and $\gamma \approx 20^{\circ}$. Wobbling bands have been identified in $^{161,163,165,167}$Lu \cite{Od01, Schoenwasser03, Bringel05, Amro03, Hagemann} and $^{167}$Ta \cite{Hartley09},
which are built on configurations that contain an odd $i_{13/2}$ proton. 
As discussed in Ref.\,\cite{Hartley11}, the  highly-aligned odd proton plays a pivotal role in generating the wobbling excitations. 
The presence of an odd $i_{13/2}$ proton drives the nuclear shape toward large deformation thereby stabilizing a TSD shape. In addition it causes a general lowering 
of the wobbling frequency. This decrease made it possible to observe the  one- and two-phonon wobbling excitations as individual bands, because
it prevented them from being immersed among the numerous particle-hole excitations.
A typical band spectrum of the odd-A wobblers is shown in Fig.\,\ref{lu163}. The wobbling frequency {\em decreases} with the spin $I$, in contrast to the 
simple wobblers shown in 
Figs.\,\ref{simpletriax} and \ref{ru112}.

Following the  discovery of the first wobbling structure in $^{163}$Lu~\cite{Od01}, the 
quasiparticle triaxial rotor (QTR) model was  used to describe the 
wobbling mode, see Refs.\,\cite{Ha02,Ha03,Ta06,Ta08}. Subsequent microscopic RPA calculations 
were able to reproduce experimental results, see Refs.\,\cite{Ma02,Ma04,Sh08,Oi07,Sh09}.  
In particular, the large ratios B(E2)$_{con}$/B(E2)$_{in}$  of 
inter-band to intra-band  E2 transitions could be 
described in both approaches.  However,  the calculated wobbling frequencies of the  QTR model with the assumptions 
of Refs.\,\cite{Ha02, Ha03} about the three
MoI distinctly disagreed with experiment. Instead of the experimentally observed decrease, the 
wobbling frequency increased with the spin $I$ (c.f  Fig.\,\ref{lu163} ). The same was found 
for  all the other  cases from the TSD region (c.f Ref.\,\cite{Hartley11}). The RPA calculations, on the other hand,
were able to reproduce the decrease of the wobbling frequency with $I$ \cite{Ma02,Ma04,Sh08,Oi07,Sh09}.

In this paper we readdress the wobbling mode in the framework of the QTR model.  
In section \ref{sec:semi} we carry out a semiclassical analysis of the QTR model assuming that the a.m. of
the odd particle is rigidly aligned with one of the principal axes. This leads to the concept of
a "transverse wobbler" and explains why its wobbling frequency decreases with $I$, in contrast to the "simple wobbler"
usually considered.  Staying within the frozen alignment approximation, simple analytical expressions for the 
energies and transition matrix elements are derived in section \ref{sec:harm}, which generalize the well known 
expressions for the simple wobbler. In section \ref{sec:TPR} we present detailed QTR calculations for  $^{163}$Lu and $^{135}$Pr.
We assume an  arrangement of the MoI with respect to the principal axes that differs from the one in the previous
QTR calculations \cite{Ha02,Ha03,Ta06,Ta08}. We shall demonstrate that this  "transverse" arrangement, which is consistent with
microscopic calculations,  results 
in  both  large  inter-band to in-band $B(E2)$ ratios and a decrease of the wobbling frequency with spin. The results account well
for the  experimental findings.

\section{Simple, transverse and  longitudinal Wobblers}\label{sec:semi}
First, we review  the semiclassical analysis of the familiar  triaxial rotor, which we
denote "simple rotor", to distinguish it from the cases to be discussed below.   
Bohr and Mottelson\,\cite{Bo75} discussed  the rotational motion  
and pointed out the possible existence of wobbling excitations at high spin.
They started from the Hamiltonian of a rigid triaxial rotor, which in the body fixed frame is given by
\begin{equation}
\label{rham}
H = A_3\hat J_3 ^2 +  A_1\hat J_1 ^{2} + A_2\hat J_2^{2}, \\
\end{equation}
where $J_{k}$ are the a.m. components and $A_{k}$ the rotational parameters with respect to the principal axes
$k=1,2,3$. Correspondingly, the moments of inertia ${\cal J}_k$ are 
given by the relation
\begin{equation}
{\cal J}_k = \frac{\hbar^2}{2 A_k}.
\end{equation}
The motion of the a.m. vector $\vec J=(J_1,J_2,J_3)$   can be conveniently visualized by considering the classical orbits of $\vec J$. These orbits  are determined by
the conservation of a.m. (a.m.)
\begin{equation}\label{amsphere}
J^2 = J_1^2 + J_2^2 + J_3^2 = I (I+1)
\end{equation}
 and energy
\begin{equation}\label{enellipsoid1}
E= A_3J_3 ^2 + A_1J_{1}^{2} + A_2J_{2}^{2}.
\end{equation} 
\begin{figure}
\hspace*{1.cm}
\vspace*{-1.cm}
\includegraphics[scale=0.7]{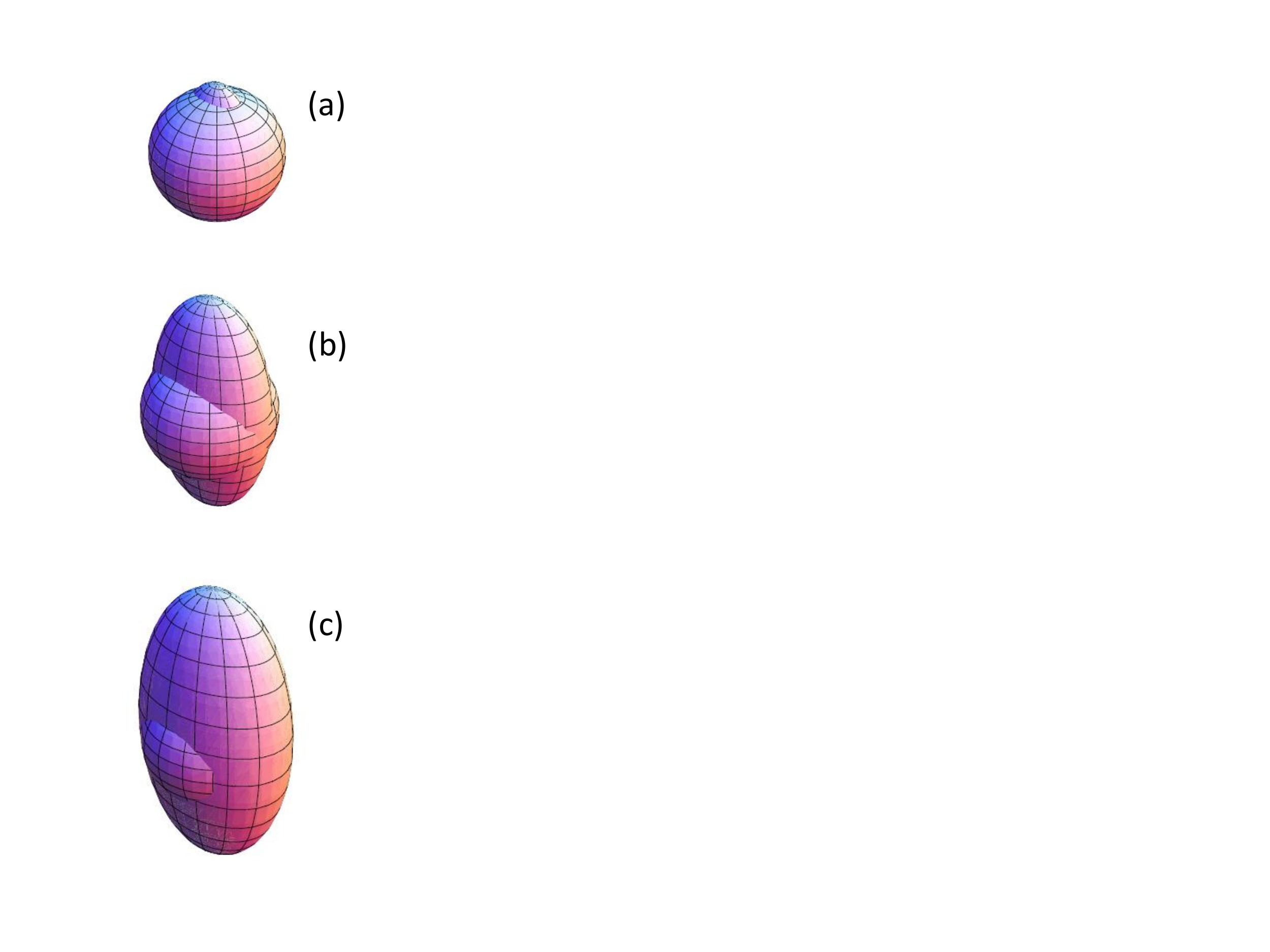}
\vspace*{-0.5cm}
\caption{\label{sintersect} (Color online) 
Classical a.m. sphere and energy ellipsoid for a simple triaxial rotor with the rotational parameters  $A_1=6A_3$ and
$A_2=3A_3$. The intersection line is the classical orbit of the a.m. vector relative to the 
body fixed frame. The three panels (a), (b) and (c) correspond to the orbits 1, 4, and 6 in Fig.\,\ref{sorb}. }
\end{figure}  

The classical orbit of $\vec J$ is the intersection of the a.m. sphere (\ref{amsphere}) with the energy
ellipsoid (\ref{enellipsoid1}). 
Let us assume that  the axes  are chosen such that  $\,A_1>A_2>A_3\,$ or accordingly ${\cal J}_1 <{\cal J}_2 <{\cal J}_3$.

Fig.\,\ref{sintersect} illustrates three types of orbits for a given a.m. value $J$, which is the radius of the a.m. sphere  (\ref{amsphere}). 
We assume for the rotational parameters  $A_1=6A_3$ and $A_2=3A_3$ and use the value $A_3$ as energy scale. 
The size of the energy ellipsoid increases with the energy $E$.  The yrast line corresponds to touching 
between the surfaces (\ref{amsphere}) and (\ref{enellipsoid1}) at  the  point $J_3=J$.   The nucleus rotates uniformly about the 3-axis 
with the maximal MoI at the energy  $E(J)=A_3 J^2$. The upper panel shows an orbit just above the yrast line, which represents the harmonic 
wobbling motion as discussed by Bohr and Mottelson. The middle panel shows the orbit called separatrix.
It has the energy of the unstable uniform rotation about the 2-axis with the intermediate MoI. 
The frequency of this orbit is zero, because it takes infinitely long time to get to or to depart from the point of the labile 
equilibrium (uniform rotation about the 2-axis).  The orbits with larger energy than the one of the separatrix  revolve the 1-axis. 
The lower panel shows one example. 

The  wobbling excitations are  small amplitude oscillations
of the a.m. vector $(J_1, J_2, J_3 )$ about the 3-axis of the largest MoI.
Their energy is given by a harmonic spectrum of wobbling quanta \cite{Bo75}
\begin{equation}
H = A_3I(I+1) + \left( n + \frac{1}{2} \right)\hbar\omega_w,
\end{equation}
where $n$ is the number of wobbling quanta and
the wobbling frequency $\hbar\omega_w$ is equal to 
\begin{equation}\label{omsrot}
\hbar\omega_w = 2I\,[(A_1 -A_3)(A_2-A_3)]^{1/2}.
\end{equation}

Quantum mechanically one has to take into account the invariance  of the rotor with respect to rotations by $\pi/2$ about
its principal axes. It has the consequence that the states 
have a signature quantum number $\alpha$=mod($I$,2)=$I$+even, which is fixed by the quantum number $I$ of the a.m.. 
In even-even nuclei the signature alternates between 0 and 1, starting with 0 for the yrast line.  
The first wobbling band ($ n$=1) has $\alpha=1$.  Above the separatrix,
there are two classical orbits with the same  energy, which revolve the positive and negative 1-half axes. 
The corresponding quantal states are symmetric and antisymmetric combinations with signature 0 and 1, respectively. 
Fig.\,\ref{simpletriax} illustrates how the separatrix divides the quantal spectrum into the two types of states that correspond to classical orbits revolving the 3- and 1-axes. 
The inset shows how the harmonic wobbling mode emerges with increasing a.m. $I$. 

In order to describe quantitatively the motion of $\vec J$ we introduce the canonical variables $J_3$ and $\phi$,
\begin{eqnarray}
J_1 =J_{\perp} \mbox{cos}\phi,~~
J_2 = J_{\perp} \mbox{sin}\phi,~~
J_{\perp} = \sqrt{J^2 - J_3^2},
\label{variables}
\end{eqnarray}
where $\phi$ is the angle of the 1-axis with the projection of $\vec J$ onto the 1-2 plane. 
Inserting the definitions (\ref{variables}) into Eqs.(\ref{amsphere},\ref{enellipsoid1}) gives the following equation for the orbits 
\begin{eqnarray}\label{phi}
\phi(J_3) = \arcsin\sqrt{\frac{E-A_1(J^2-J_3^2)-A_3 J_3^2}{(A_2-A_1)(J^2-J_3^2)}}.
\end{eqnarray}

\begin{figure}
\vspace{-0.5cm}
\includegraphics[scale=0.33] {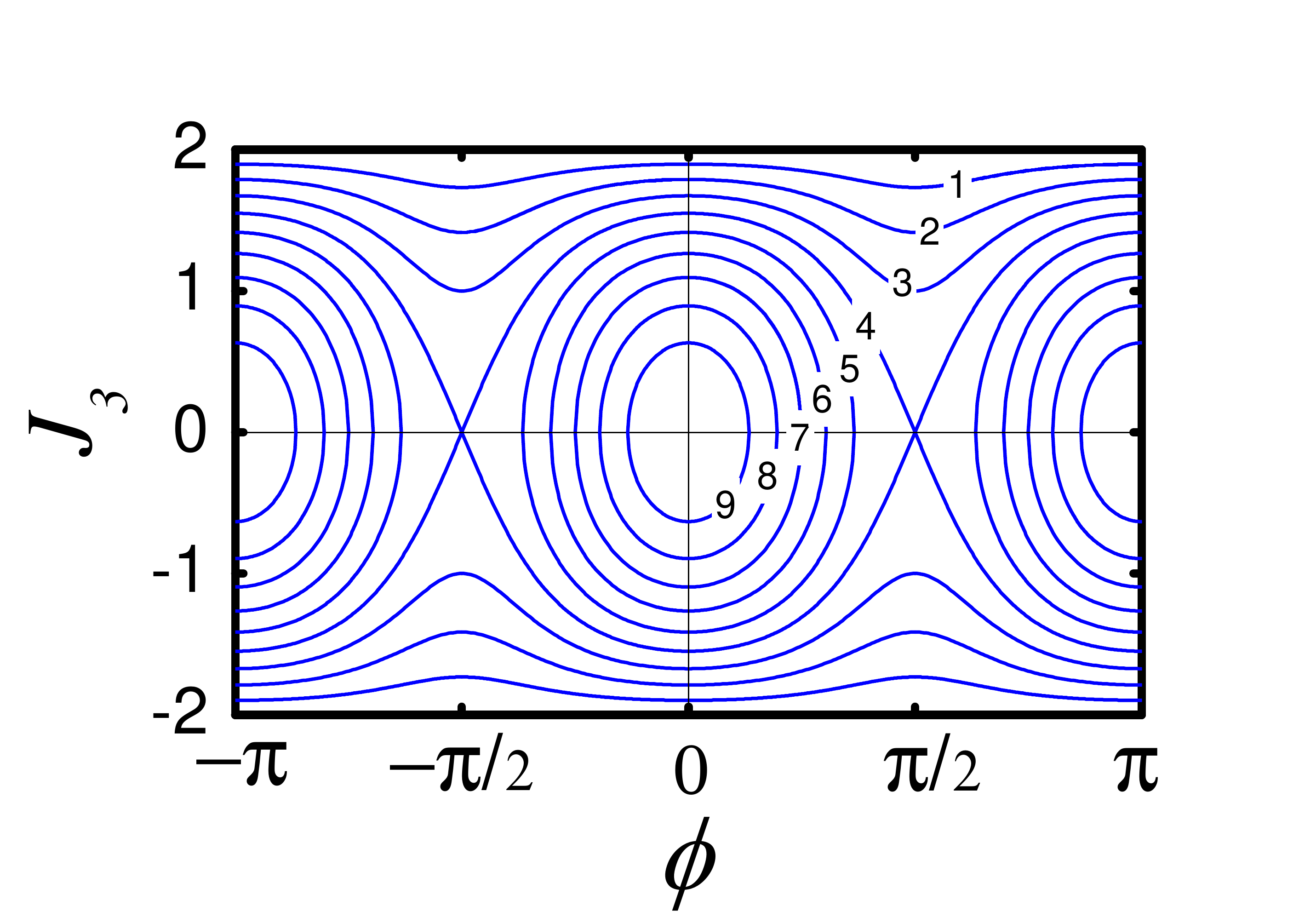}    
\caption{\label{sorb} (Color online) 
Classical orbits (blue lines) of the a.m. vector for a simple triaxial rotor  with the rotational  parameters  $A_1=6A_3$ and
$A_2=3A_3$. The a.m. is  $J=2$.
The series of the orbits 1-9 corresponds to the energies $E = 6,8,\dots,22$ 
in terms of the energy  unit $A_3$.}
\end{figure}

The phase space for the one-dimensional
motion on the a.m. sphere is $-\pi \leq \phi \leq \pi$ and $-J\leq J_3 \leq J$. Fig.\,\ref{sorb} shows a series of orbits in the 
phase space.  The stable minimum lies 
at the point where  $J_3 = J$ (here equal $2\,\hbar\,$). 
 Below, for a better comparison with the cases when a particle is present, we use the particle a.m. $j$ as the unit for
the total a.m..    Orbit 1 corresponds to wobbling about the 3-axis. Orbit 4 is the separatrix. Orbit  7 corresponds to
wobbling about the 1-axis. According to classical mechanics, the period of the orbit is 
$T=2\pi dS/dE$, where $S$ is the phase 
space area enclosed by the orbit. The orbits in Fig.\,\ref{sorb} are calculated for an equidistant  set of energies.
As seen, the difference $\Delta S$ is maximal near the separatrix, which means there the period $T$ has a maximum and the 
frequency $\omega=2 \pi/T$ has a minimum. In classical mechanics the energy increases continuously, and the frequency 
of the separatrix goes to zero, as mentioned above. In quantum mechanics the increase of the energy is discrete, such that 
   $\Delta S=2\pi\hbar$, i.e. the energy distance between adjacent levels has a minimum at the separatrix.   
For comparison we present in Fig.\,\ref{simpletriax} the complete series of quantal band structures
as calculated for the triaxial rotor with the same rotational parameters as in 
Fig.\,\ref{sorb}. 

Now we discuss  the wobbling excitations in odd-A triaxial nuclei.
In order to account for the presence of a high-j odd quasiparticle, the triaxial rotor Hamiltonian 
must be replaced by the   quasiparticle triaxial rotor (QTR) Hamiltonian
\begin{equation}
\label{prham}
H = h_{dqp}+A_3(\hat J_3 - \hat j_3)^2 +  A_1(\hat J_1 - \hat j_1)^{2} + A_2(\hat J_{2}-\hat j_2)^{2}, 
\end{equation}
where $\hat j_k$ is the a.m. of the odd quasiparticle and $\hat J_k$ the total a.m. The term $h_{dqp}$ describes the coupling of
the odd quasiparticle to the triaxial core.  Qualitatively, the coupling aligns the $\vec j$ of a high-j particle
with the short $(s)$ axis, because this orientation corresponds to maximal overlap between the density distribution of the particle
and the triaxial core, which minimizes the attractive short range core-particle interaction. Likewise, the coupling aligns  
 the $\vec j$ of a high-j hole with the long $(l)$ axis,  because this orientation corresponds to minimal overlap between the density distribution of the hole
and the triaxial core, which minimizes the repulsive short range core-hole interaction. The coupling aligns the $\vec j$ of a quasiparticle from
a half-filled high-j orbital with the medium $(m)$ axis.  These coupling schemes can be verified by microscopic calculation within the frame of the cranking model.

\begin{figure}
\vspace*{-0.5cm}
\includegraphics[scale=0.33]{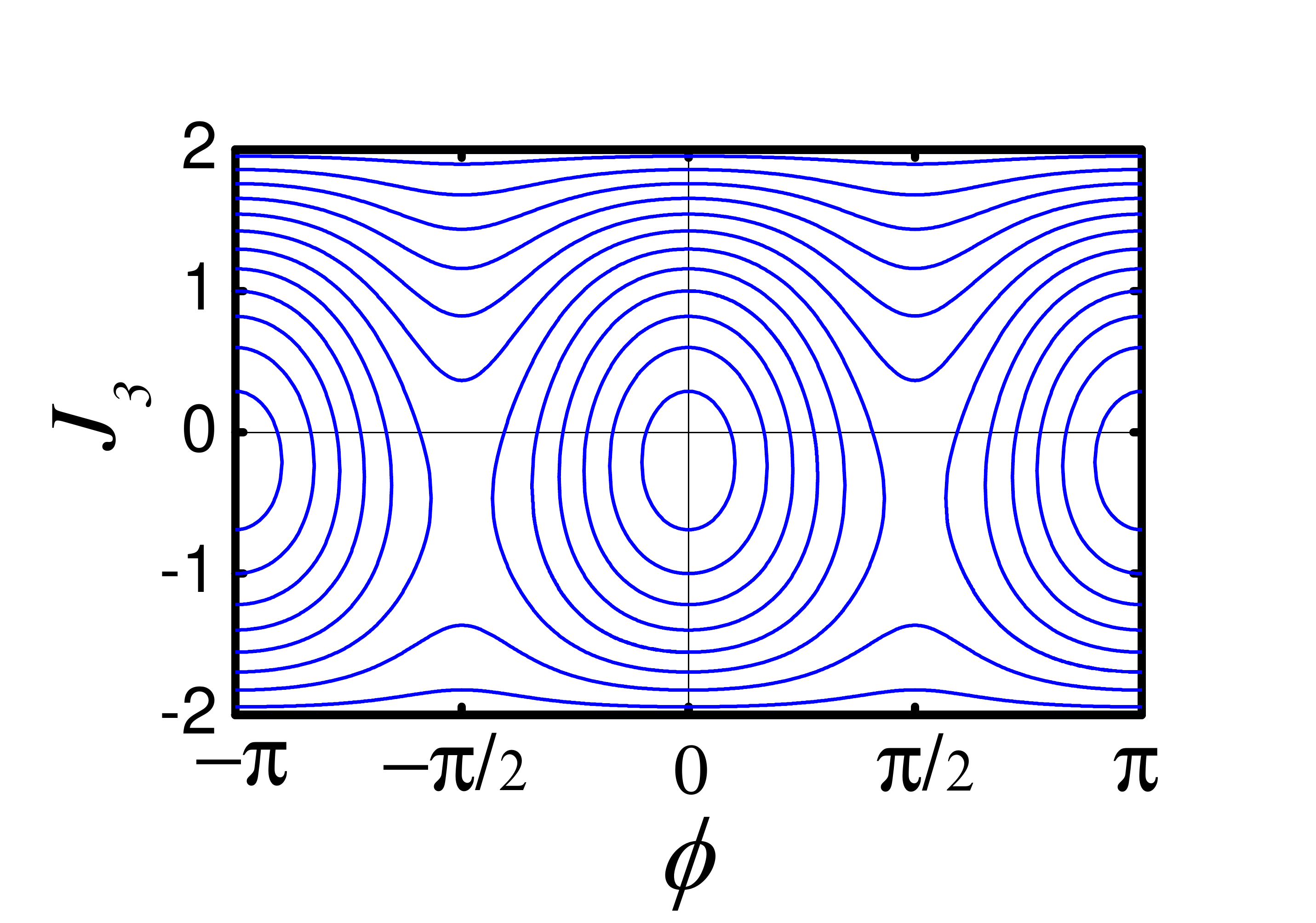}  
\caption{\label{lorb} (Color online) 
Classical orbits (blue lines) of the a.m. vector for a longitudinal triaxial rotor with a high-j particle. The rotational  parameters are   $A_1=6A_3$ and
$A_2=3A_3$. The a.m. is $J=2$ in units of the particle a.m. $j$. The topmost orbital has $E=2$ in terms of the energy unit $A_3j^2$ and 
beneath the energy is increasing in steps of 2.} 
\end{figure}  

The coupling  of the high-j quasiparticle to the triaxial rotor considerably modifies  
the motion of the a.m. vector $\vec J$ with respect to the body fixed frame.
To carry out the semiclassical analysis
we assume that the a.m. of the odd quasiparticle is rigidly aligned with
one of the principal axes of the triaxial rotor. This
  "Frozen Alignment" (FA) approximation idealizes the above discussed tendency of the quasiparticle to align its
 a.m. with one of the principal axes according to its particle-hole character.
In the following we assume that the alignment is along the 3-axis. Again, 
the motion of the a.m. vector is visualized by the classical orbits of $\vec J$, which are determined by
the conservation of  a.m. and the energy.
The classical orbits of $\vec J$ are the intersection of  the a.m. sphere\,(\ref{amsphere}) with the shifted energy
ellipsoid
\begin{equation}\label{enellipsoid}
E= A_3(J_3 - j)^2 + A_1J_{1}^{2} + A_2J_{2}^{2}.
\end{equation}  
Accordingly, in Eq.(\,\ref{phi}) for the orbit $\phi(J_3)$ the term $A_3 J_3^2$ has to be replaced by the shifted term $A_3(J_3-j)^2$.
\begin{figure}
\includegraphics[scale=0.33]{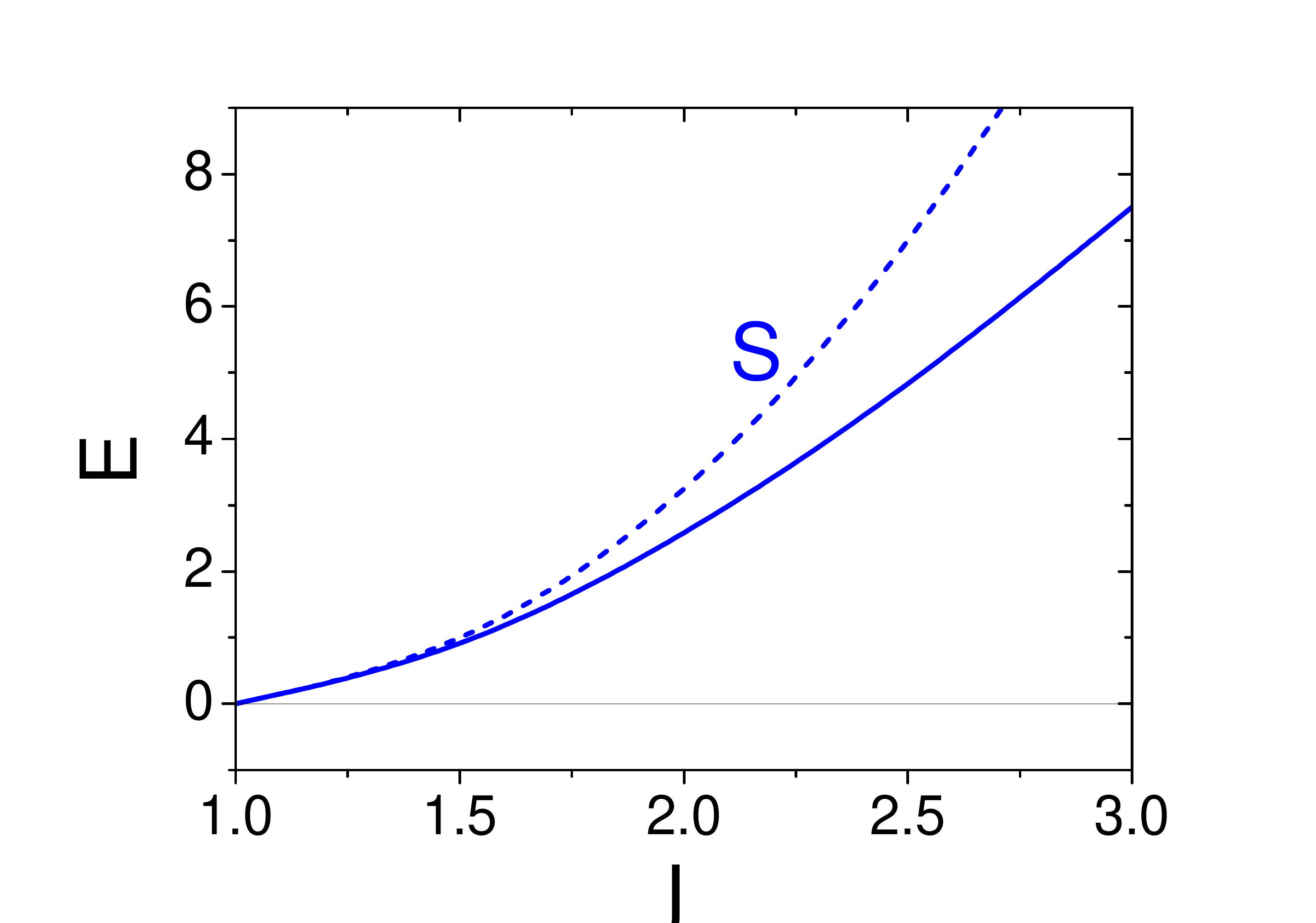}  
\caption{\label{tEJ}
(Color online) Energy E of the yrast line and the separatrix (S) for the transverse rotor with the rotational  parameters  $A_1=6A_2$ and
$A_3=3A_2$. The unit of the a.m. J is  $j$, and the energy unit is $A_2j^2$.}
\end{figure}  

As discussed above, the triaxial shape of the rotor determines the orientation of quasiparticle with respect to its principal axes. However,
it also determines the ratios between the three MoI, which are of the hydrodynamic type: The MoI of the medium ($m$) axis
is always the largest. This can be inferred   from a simple argument that holds for both the hydrodynamic  and quantal systems. The
MoI is zero for rotation about a symmetry axis and increases with the deviation from axial symmetry of the  axis. 
The triaxial shape deviates most strongly from axial symmetry with respect to the $m$-axis, which results  in the largest 
MoI. Microscopic   calculation based on the cranking model typically give the order   $ {\cal J}_l <{\cal J}_s <{\cal J}_m$
using again the notation $l,s$ and $m$ for the long, short and medium axis , respectively. The microscopic ratios
deviate from the hydrodynamic ones. In particular,  for $\gamma=30^\circ$ one still has  $ {\cal J}_l <{\cal J}_s <{\cal J}_m$
in contrast to the hydrodynamic ratios $ {\cal J}_l ={\cal J}_s <{\cal J}_m$ (see Tab. \ref{tab:parameter} below). 

One must distinguish between the quasiparticle a.m. vector $\vec j$ 
being aligned with the $m$-axis with the largest MoI, which we refer to as the $longitudinal$ case, 
and the vector  $\vec j$ being perpendicular to the $m$-axis, which we refer to as the $transverse$ case. 
A quasiparticle with predominantly particle character, which 
emerges from the bottom of a deformed j-shell,   aligns its  $\vec j$  with the $s$-axis. It combines with the triaxial rotor (TR)  to a 
transverse QTR system.  
A quasiparticle with predominantly hole character, which 
emerges from the top of a deformed j-shell,   aligns its  $\vec j$  with the $l$-axis. It  couples with the 
TR to a transverse QTR too.   A quasiparticle, which emerges
from the middle of the j-shell and tends to align with the $m$-axis, couples with the TR to a longitudinal QTR. 
The Coriolis force tends to realign  $\vec j$ from the $s$- or $l$- axes toward the $m$-axis. It may overcome
the coupling to the deformed potential, resulting in a change from the transverse to the longitudinal  mode.
 
The longitudinal QTR is similar to the simple rotor, only 
that the energy ellipsoid, Eq.(\ref{enellipsoid}), is shifted upwards by $j$.  
The yrast line corresponds to uniform rotation about the 3-axis and the 
lowest excited states represent wobbling about this axis as shown in the upper panel of Fig.\,\ref{sintersect}. 
The orbits in phase space are shown in Fig.\,\ref{lorb}.
The wobbling bands in this odd-A case have alternating signature $\alpha=\pm$ \,mod($j$,2), starting with 
 $\alpha=$\,mod($J$,2) at the yrast line. The wobbling frequency increases with a.m.. 
 We mention that the "reversed arrangement" (as compared to the hydrodynamic one) of the MoI used in Refs.\,\cite{Ha02, Ha03}  
 for the description of wobbling bands in the Lu isotopes corresponds to a longitudinal QTR. 
This arrangement is inconsistent with above discussed natural order  $ {\cal J}_l <{\cal J}_s <{\cal J}_m$, because  
the odd i$_{13/2}$ quasiproton has particle character and as such aligns  its $\vec j$ with the $s$-axis, to which 
 the maximal MoI is assigned.

\begin{figure}
\vspace*{0.5cm}\includegraphics[scale=0.15]{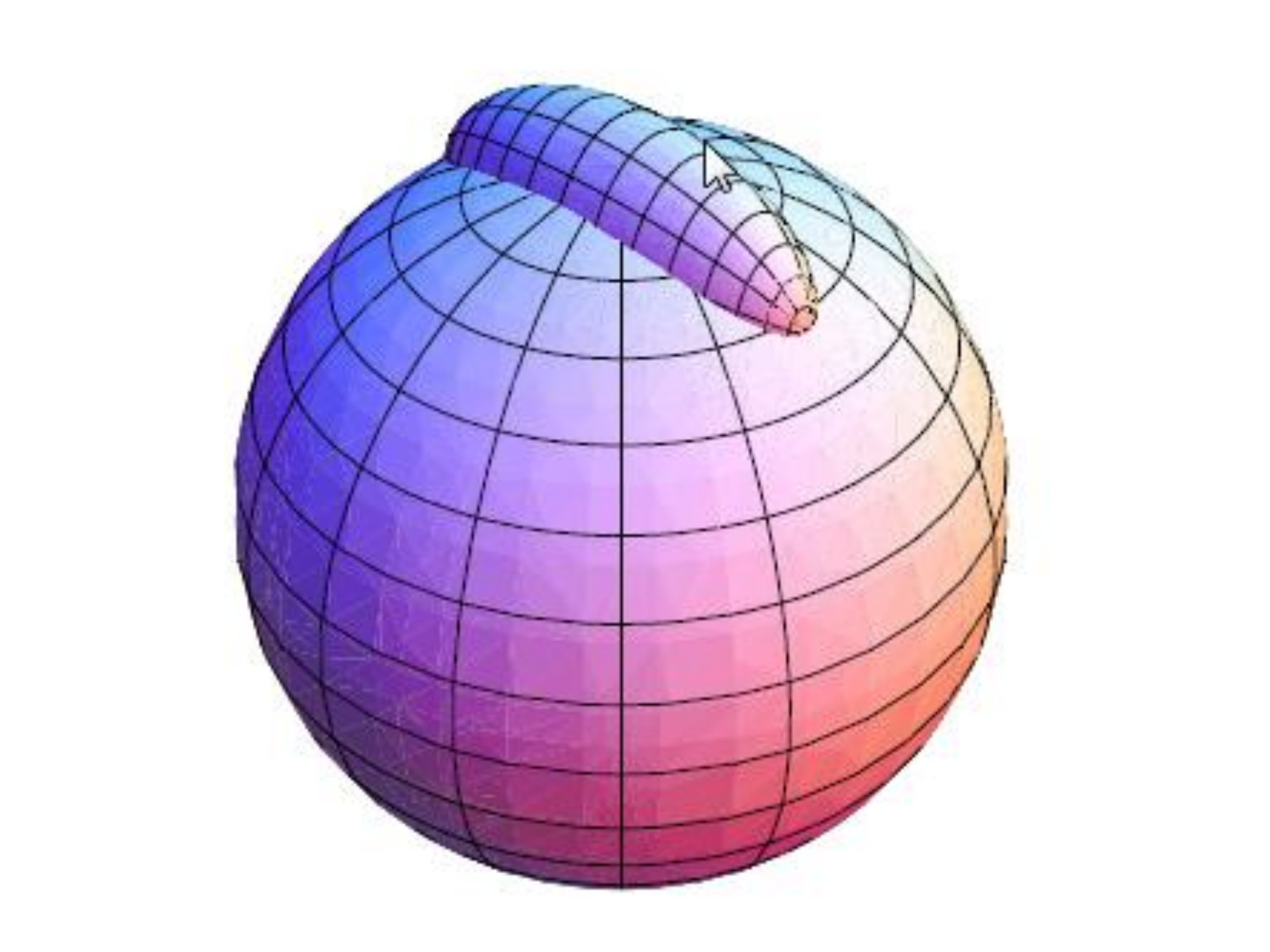}
\caption{\label{tintersectl} (Color online)
Angular momentum sphere and energy ellipsoid for a transverse QTR with the rotational  parameters  $A_1=6A_2$ and
$A_3=3A_2$. The intersection line is the classical orbit of the 
a.m. vector relative to the 
body fixed frame. This line corresponds to the lowest energy orbit  in Fig.\,\ref{torbl}. }
\end{figure}

\begin{figure}
\hspace*{1.5cm}
\includegraphics[scale=0.8]{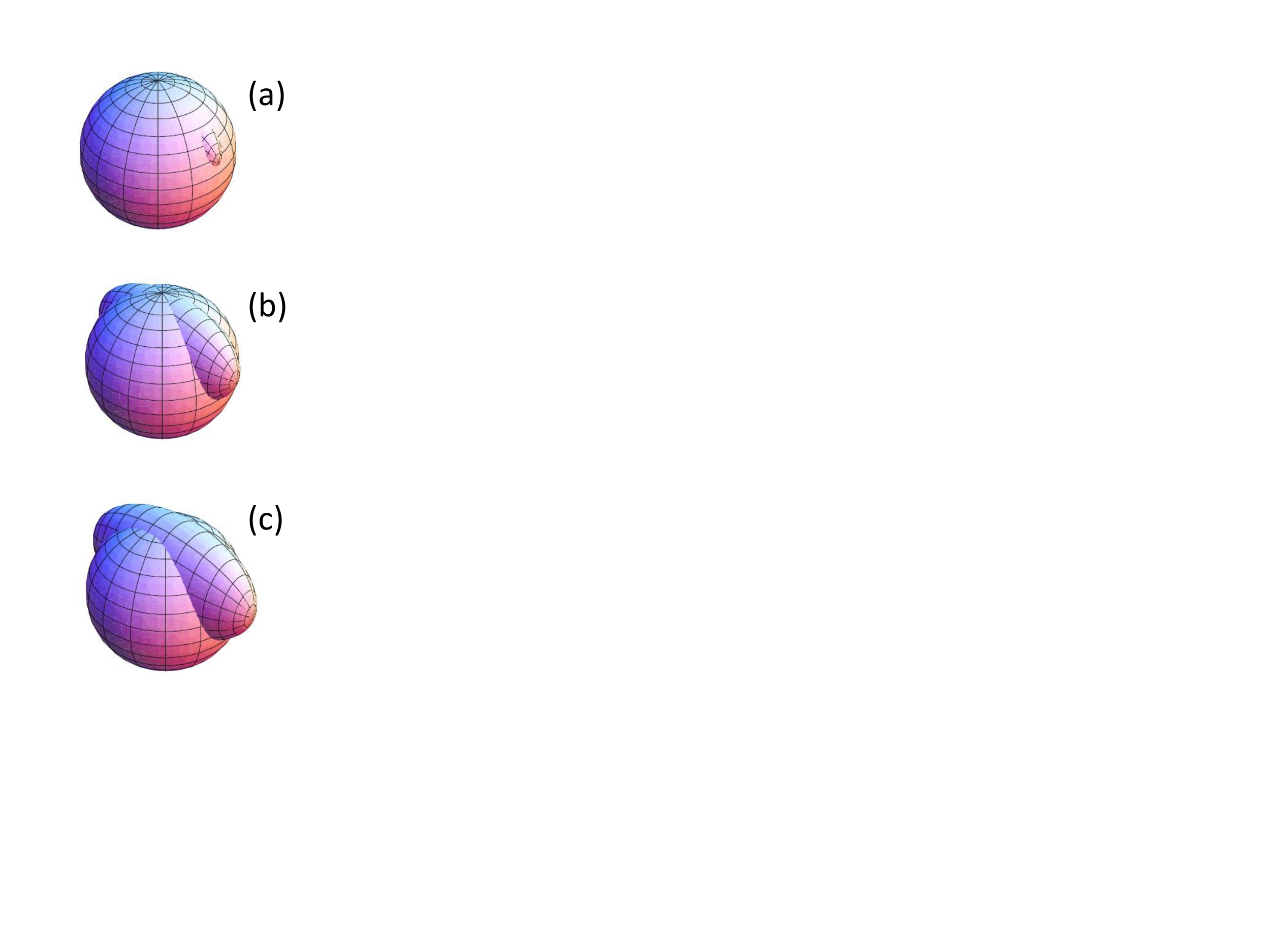}
\hspace*{1.5cm}
\vspace*{-4.cm}
\caption{\label{tintersecth} (Color online)
Angular momentum sphere and energy ellipsoid for a transverse QTR with the rotational  parameters  $A_1=6A_2$ and
$A_3=3A_2$. The intersection line is the classical orbit of the a.m. vector relative to the body fixed frame. The three panels (a), (b) and (c) correspond to the orbits with the smallest energy, the separatrix, and the next orbit with higher energy, respectively,  in accordance with the orbits 
shown in Fig.\,\ref{troth}.  }
\end{figure}

\begin{figure}
\includegraphics[scale=0.33]{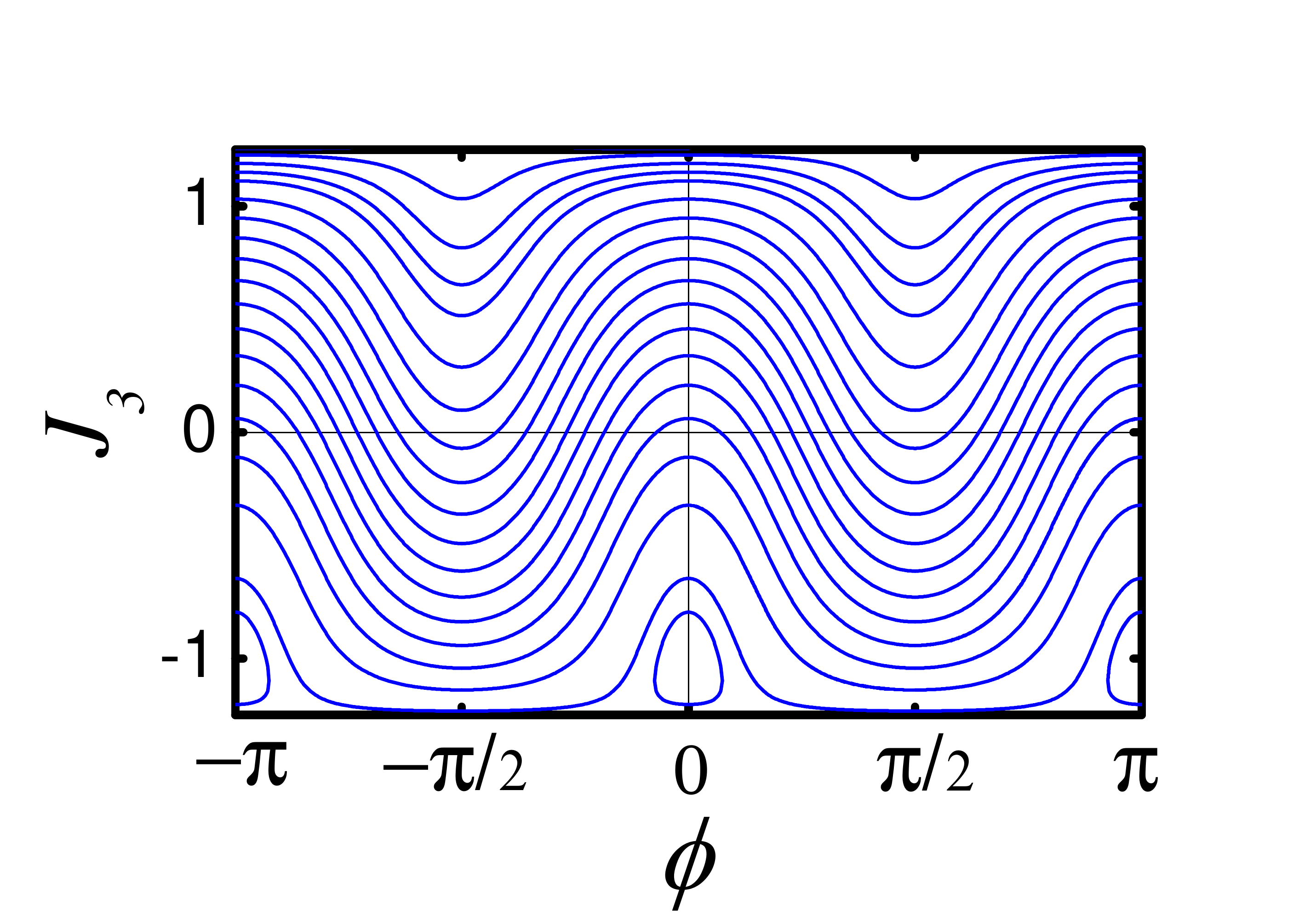}  
\caption{\label{torbl} (Color online) 
Classical orbits (blue lines) of the a.m. vector for a transverse triaxial rotor with the rotational  parameters  $A_1=6A_2$ and
$A_3=3A_2$. The a.m. is $J=1.25j$ being below the critical a.m. $J_c$. The a.m. unit is  $j$. The energy increases from the top to the bottom.
The energy difference between the orbits is 1, where the energy unit is $A_2j^2$. The separatrix is located in the lower part of the figure right above the closed orbits at $\phi=0,\pm\pi $. }
\end{figure}

\begin{figure}
\includegraphics[scale=0.33]{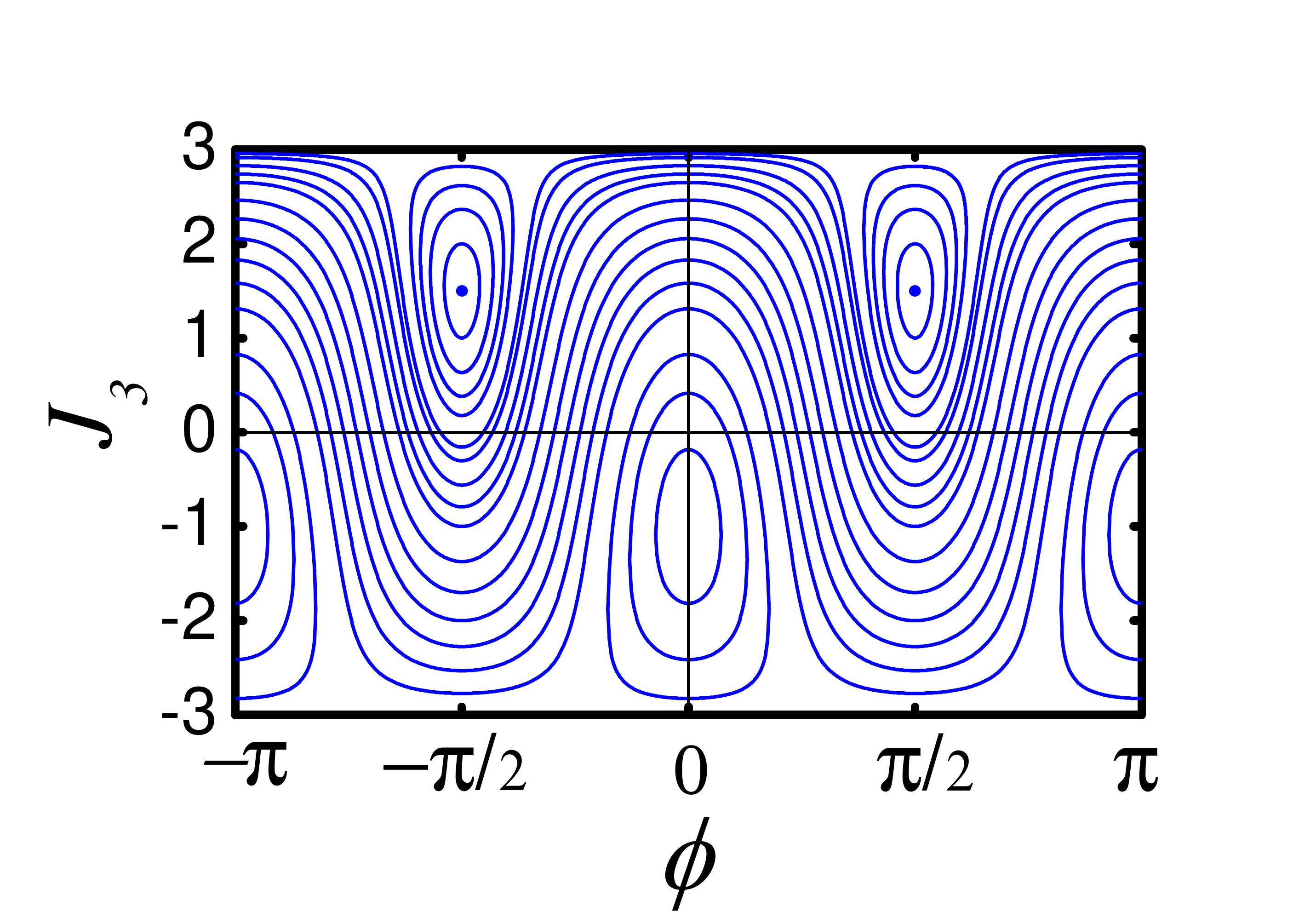}   
\caption{\label{troth} (Color online) 
Classical orbits (blue lines) of the a.m. vector for a transverse triaxial rotor with the rotational  parameters  $A_1=6A_2$ and
$A_3=3A_2$. The a.m. is $ J=3$ which is 
above the critical a.m. $J_c$. The a.m. unit is  $j$.
The energy difference between the orbits is 4, where the energy unit is $A_2j^2$. The dot indicates the yrast "orbit", from
where the energy increases.  }
\end{figure}

 For the analysis of the transverse QTP, we assume the quasiparticle to be  particle-like, i.e. the 3-axis is the $s$-axis. Further, we 
 assign the 2-axis to the $m$-axis with the largest 
 MoI and the 1-axis to the $l$-axis with the smallest MoI. 
 (The axes 2 and 3 are exchanged 
 compared to the discussion of the simple TR and the longitudinal QTR.)  The pertinent figures are generated with the rotational parameters $A_1=6A_2$, $A_2=A_2$ and $A_3=3A_2$. 
As illustrated by Fig.\,\ref{tEJ}, the resulting yrast line consists of two pieces. At low a.m. it corresponds to 
rotation about the 3-axis. The energy ellipsoid touches the a.m. sphere at the point $J_3=J$ on the 3-axis and the yrast energy is $E=A_3(J-j)^2$.
The low energy  orbits above yrast represent  wobbling about the  3-axis.   Fig.\,\ref{tintersectl} displays the intersection line of
 the lowest orbit in Fig.\,\ref{torbl}, which shows the orbits in phase space. At the critical a.m. 
 $J_c=jA_3/(A_3-A_2)$ the rotational axis of the yrast line flips to the direction of the point 
 $J_1=0$,   $J_2=\sqrt{J^2-J_c^2}$, $J_3=J_c$, where the energy ellipsoid touches the a.m. sphere from inside. This means,  the  axis  of 
 uniform rotation is tilted   into the $s$-$m$-plane. Quantum mechanically, rotation about such a tilted axis corresponds to merging of the two
 signatures into a $\Delta I=1$ band.
The upper panel of Fig.\,\ref{tintersecth} displays the intersection of the energy ellipsoid with a.m. sphere for a slightly 
higher energy, which is   the first orbit enclosing the touch point ($\phi=\pi/2,\, J_3=3/2j$) in Fig.\,\ref{troth}. 

 The separatrix, which is illustrated in the middle panel of Fig.\,\ref{tintersecth}, lies at  $E=A_3(J-j)^2$.
Above the separatrix  the orbits revolve the 
3-axis as shown in the lower panel of Fig.\,\ref{tintersecth}. 
For $J<J_c$ the yrast line is $E=A_3(J-j)^2$. It continues as separatrix for  $J>J_c$.  
 As discussed above for the simple rotor, the classical frequency of the separatrix is zero. 
 Analogously for the transverse QTR the frequency of the small-amplitude wobbling goes to zero at  $J=J_c$,
 where uniform rotation about the 3-axis becomes unstable, and the new branch of the yrast line starts.
 Quantum mechanically, the yrast states have signature $\alpha$=\,mod($j$,2) for $J<J_c$, and the first wobbling state has opposite signature $-\alpha$.
  It encloses the fixed area $2\pi\hbar$ in phase space, which means its energy
 decreases with $J$.    It merges with the yrast line, which becomes a $\Delta I =1$ sequence for $J>J_c$.
 In case of the longitudinal rotor, there is no bifurcation of the yrast line, which is reflected by a continuous increase of the wobbling frequency with $J$.

\section{Harmonic wobbling Model}\label{sec:harm} 
Now we consider small amplitude wobbling vibrations about the 3-axis.   
We retain the FA approximation, i.e.  the a.m. of the odd quasiparticle
is assumed to be firmly aligned with the 3-axis and can be considered as a number. Then the QTR  Hamiltonian becomes
\begin{equation}
\label{faham}
H = A_3(\hat J_3 - j)^2 + A_1 \hat J_1 ^{2} + A_2\hat J_{2}^{2}, 
\end{equation}
where $j$ is a number. We use the second order expansion  
\begin{equation}
\hat J_3=\sqrt{J^2-\hat J_1^2-\hat J^2_2}\,\,\approx J-\frac{1}{2}\left(\frac{\hat J^2_1}{J}+\frac{\hat J^2_2}{J}\right) 
\end{equation}
where $J=\sqrt{I(I+1)}$. The Hamiltonian becomes in harmonic FA (HFA) approximation 
\begin{eqnarray}\label{woham}
H = A_3( J - j)^2 +
(A_1-\bar A_3)\hat J_1 ^{2} +(A_2-\bar A_3) \hat J_2 ^{2},\\
\text{where}\quad\quad \bar A_3=A_3(J)= A_3\left(1-\frac{j}{J}\right).
\end{eqnarray}

This Hamiltonian has the form of the simple TR Hamiltonian, except that $A_3$ is replaced by the $J$-dependent
rotational  parameter $\bar A_3(J)= A_3(1-j/J)$. Therefore, one can carry over 
the expressions given by Bohr and Mottelson \cite{Bo75} replacing  
$A_3$ by $\bar A_3(J)$. \footnote{Ref.\,\cite{Bo75} uses the arrangement ${\cal J}_1 >{\cal J}_2 >{\cal J}_3$, which differs from the arrangement in this paper.
In taking over the expression one has to relabel the axes accordingly.} 
 The wobbling frequency becomes
\begin{eqnarray}\label{om1wob}
\hbar\omega_w = 2J[(A_1 -\bar A_3(J))(A_2-\bar A_3(J))]^{1/2}.
\end{eqnarray}
\noindent
Correspondingly, the E2-transition probabilities are \cite{Bo75}
\begin{eqnarray}
B(E2,~n,~I \rightarrow n, I\pm 2)=\frac{5}{16\pi}e^2 Q^2_2,\\
B(E2,~n,~I \rightarrow n-1, I-1)=&&\nonumber \\
\quad\quad\quad\frac{5}{16\pi}e^2\frac{n}{J}(\sqrt{3} Q_0x-\sqrt{2}Q_2y)^2,\\
B(E2,~n,~I \rightarrow n+1, I+1)=&&\nonumber \\
\frac{5}{16\pi}e^2\frac{n+1}{J}(\sqrt{3} Q_0y-\sqrt{2}Q_2x)^2,
\end{eqnarray}
where
\begin{eqnarray}
\genfrac{[}{]}{0pt}{}{x}{y}=
\genfrac{(}{)}{0pt}{}{1}{-\text{sign}(\beta)}
\left[\frac{1}{2}\left(\frac{\alpha}{\,\,\hbar\omega_w}\pm1\right)\right]^{1/2},\quad\quad\\
\alpha=\left(A_1+A_2-2\bar A_3(J)\right)J,~~\quad\beta=\left(A_1-A_2\right)J
\quad\quad 
\end{eqnarray}
and $Q_0$ and $Q_2$ are the quadrupole moments of the triaxial charge density relative to the 3-axis.
The transition probabilities B(M1) can be derived analogously to the B(E2) in Ref.\,\cite{Bo75} by assuming that
the aligned quasiparticle generates a magnetic moment component $\mu_3=(g_j-g_R)j$.
One finds
\begin{eqnarray}\label{hfa-bm1}
B(M1,n,I \rightarrow n-1, I-1)=\frac{3}{4\pi}\frac{n}{J}[j(g_j-g_R)x]^2,\,\,\,\\
B(M1,n,I \rightarrow n+1, I+1)=\frac{3}{4\pi}\frac{n+1}{J}[j(g_j-g_R)y]^2.\,\,\,
\end{eqnarray}
The harmonic approximation is valid as long as \mbox{$J_1^2+J_2^2<<J^2$}. For  the first
wobbling excitation this leads to the condition \cite{Bo75}
\begin{eqnarray}\label{validity}
\frac{3(A_1+A_2-2\bar A_3)}{2(A_1-\bar A_3)^{1/2}(A_2-\bar A_3)^{1/2}}<<J. 
\end{eqnarray}    

Let us discuss the wobbling energy, Eq.\,(\ref{om1wob}), in more detail. It can be rewritten as
\begin{multline}\label{om2wob}
\hbar\omega_w =\\
\frac{j}{{\cal J}_3} \left[ \left( 1 + \frac{J}{j}\left( \frac{{\cal J}_3}{{\cal J}_1} - 1\right) \right) 
\left( 1 + \frac{J}{j}\left( \frac{{\cal J}_3}{{\cal J}_2} - 1 \right) \right) \right]^{1/2}.
\end{multline}

For the longitudinal QTR, the relation ${\cal J}_3 > {\cal J}_1,{\cal J}_2$ holds.  Both bracketed terms in in Eq. (\ref{om1wob}) are positive, and the 
wobbling frequency increases with a.m.. 

In case of the transverse QTR with an odd particle one has ${\cal J}_3 > {\cal J}_1$ but ${\cal J}_3 <{\cal J}_2$.
Then,  the factor $1+J({\cal J}_3/{\cal J}_2-1)/j$ in  Eq.~(\ref{om2wob}) decreases with $J$, and
 the wobbling energy will also decrease  for sufficiently large $J$. It reaches 
zero at $J_c=j{\cal J}_2/({\cal J}_2-{\cal J}_3$),
 which is the previously discussed
critical a.m. where the separatrix bifurcates from the yrast line.  There the yrast and the wobbling bands
 will merge into a single $\Delta I = 1$ sequence, reflecting  the fact that the rotational axis is tilted into the $s$-$m$-plane. 
  Figs. \ref{lu163TPRresult} and \ref{pr135TPRresult}  show examples. The initial increase of the wobbling frequency in the case of $^{163}$Lu is caused by the 
 factor $1+J({\cal J}_3/{\cal J}_1-1)/j$  in  Eq.~(\ref{om2wob}), which increases with $J$. It is characteristic for a situation when ${\cal J}_m $ is only slightly
 larger than   ${\cal J}_s$ but both are much larger than ${\cal J}_l$.
 
 The transverse QTR with an odd hole has typically ${\cal J}_3 < {\cal J}_1,{\cal J}_2$. Then,  both factors $1+J({\cal J}_3/{\cal J}_2-1)/j$ 
 and $1+J({\cal J}_3/{\cal J}_1-1)/j$ in  Eq.~(\ref{om2wob}) decrease with $J$, and
 the wobbling energy will always decrease with $J$. 

 The prerequisite  (\ref{validity}) for the small  amplitude approximation is violated near the instability of the transverse QTR, which means that HFA cannot be
 applied there. The assumption of frozen alignment will become invalid at a certain a.m., when the inertial forces overcome
 the coupling of the odd quasiparticle to the TR. Then the quasiparticle $\vec j$ will realign with the $m$-axis and the QTR will change from the transverse to the longitudinal mode. 
 Nevertheless, the set of equations (\ref{om1wob}-\ref{hfa-bm1}) obtained from the HFA model provides an easily manageable tool for investigating
  the properties of the QTR system. In particular, it is the analytic form of these relations which allows us to qualitatively interpret the results
  of experiment and more sophisticated calculations. 

The previous studies of the Lu isotopes with the QTR model \cite{Ha02,Ha03,Ta06,Ta08} were based on the assumption   
$A_3 < A_1,A_2$ (${\cal J}_3 > {\cal J}_1,{\cal J}_2$)
together with the quasiparticle alignment $\vec j$ along the 3-axis, which in our terminology is the longitudinal wobbler. 
Thus, the energies of the excited wobbling bands relative to the lowest 
$\pi i_{13/2}$ band must increase  with spin, which is  the opposite  trend observed in Fig.\,\ref{lu163}.  
We suggest that the observed wobbling excitations are of the transverse type, i.e. we adopt the arrangement 
 $A_2 < A_3 < A_1$, which corresponds to  the natural order  ${\cal J}_m >{\cal J}_s >{\cal J}_l$ obtained
in microscopic cranking calculations.  

Tanabe and Sugawara-Tanabe \cite{Ta06,Ta08} considered a less restrictive approximation to the QTR than HFA.
It assumes that the odd particle 
is not rigidly fixed to the core. Its a.m. may execute small amplitude oscillations.  
 They find  a moderate  coupling between the two oscillators, such
that the lowest states may be classified as being predominantly a wobbling mode of the core or a vibrational 
excitation of the odd particle. Our discussion above concerns only the first type, the wobbling modes. 
Naturally, their approach better reproduces  the exact QTR results than our HFA, which however comes at the expense of  
rather complicated expressions.

\section{QTR calculations}\label{sec:TPR}
\subsection{Model parameters}

Below we present the results of our calculations for the transverse wobblers  $^{135}$Pr and $^{163}$Lu obtained by means of the QTR model.
The results will be compared with with the HFA   and the experiment. 
Our QTR calculations have been carried out by using the core-quasiparticle-coupling  (CQM) formalism \cite{CPM}, which at variance with usual QTR 
 Hamiltonian (\ref{prham})
 is formulated in the laboratory frame of reference.  The CQM Hamiltonian is given by
\begin{figure}
\includegraphics[clip,height=6.5cm]{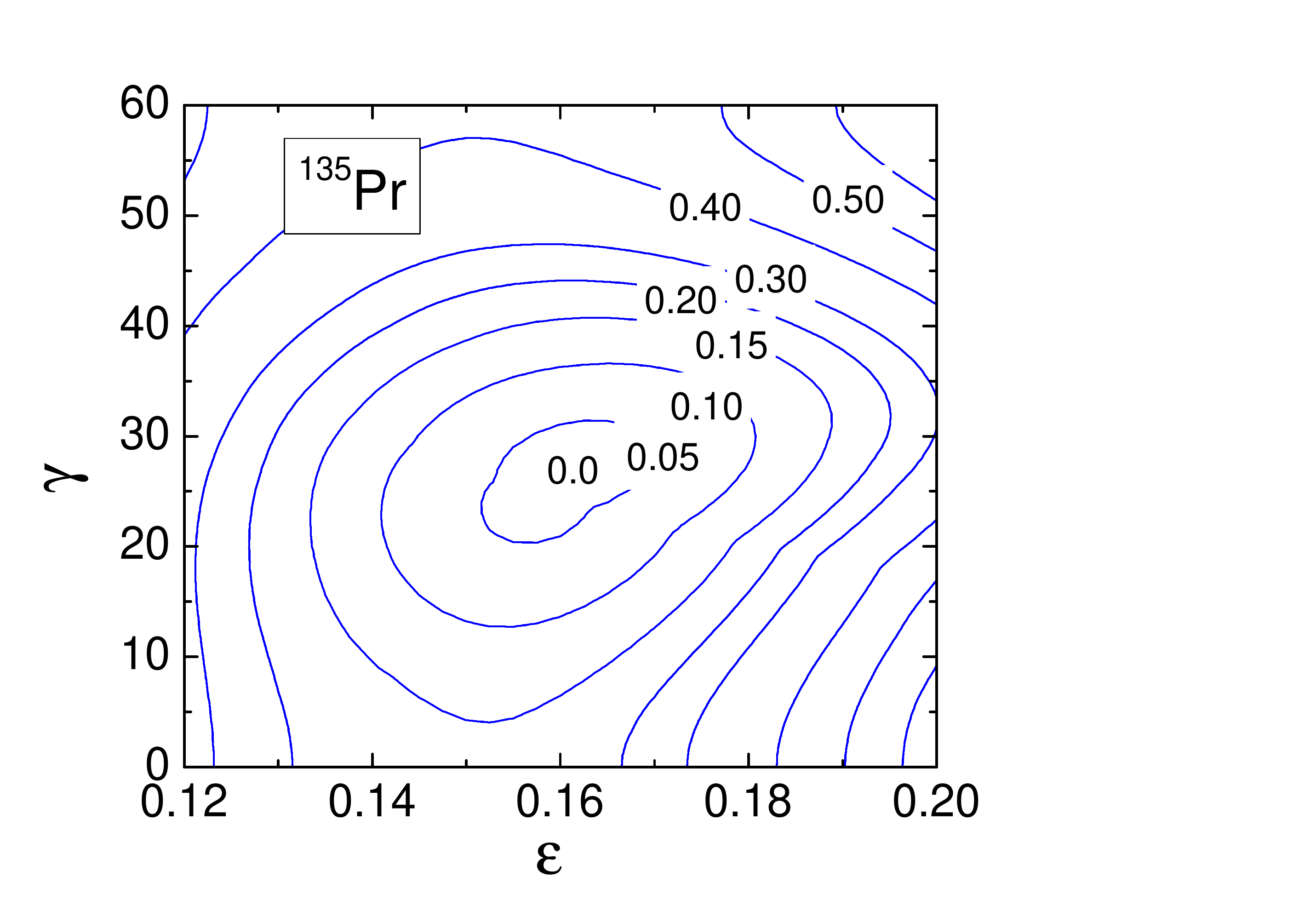} 
\hspace{-0.4cm} 
\caption{\label{TRS135Pr} (Color online)
Potential energy surface $E(\varepsilon,\gamma)$ of $^{135}$Pr calculated with the Strutinsky micro-macro method using the tilted axis cranking model \cite{TAC}. The energies attached to  the equipotential lines are in MeV. 
}
\end{figure} 

\begin{table}
\caption{ \label{tab:parameter}  Deformation parameters and moments of inertia  (in $\hbar^2$/MeV) used in the QTR and HFA calculations.
The letters $m, s, l$ denote the medium, short, long axes of the triaxial potential and charge distribution.}
 \begin{ruledtabular}
\begin{tabular}{rrccccc}
nucleus     &$\varepsilon$ &$\gamma (deg)$&model
&${\cal J}_m$&${\cal J}_s$&${\cal J}_l$  \\
\hline
135Pr & 0.16& 26 & \quad fit&21& 13 & 4 \\
     &0.16 & 26& hydrodyn   & 20 & 6   & 4\\
     &0.16 & 26& cranking &17&7&3     \\
163Lu & 0.4& 20 & \quad fit&64& 56 & 13 \\
      & 0.4 & 20& hydrodyn &  68&  29&   8  \\
      & 0.4 & 20 & cranking  & 59 &  51&  13    \\
\end{tabular}
\end{ruledtabular}
\end{table}

\begin{figure}
\includegraphics[clip,height=6.5cm]{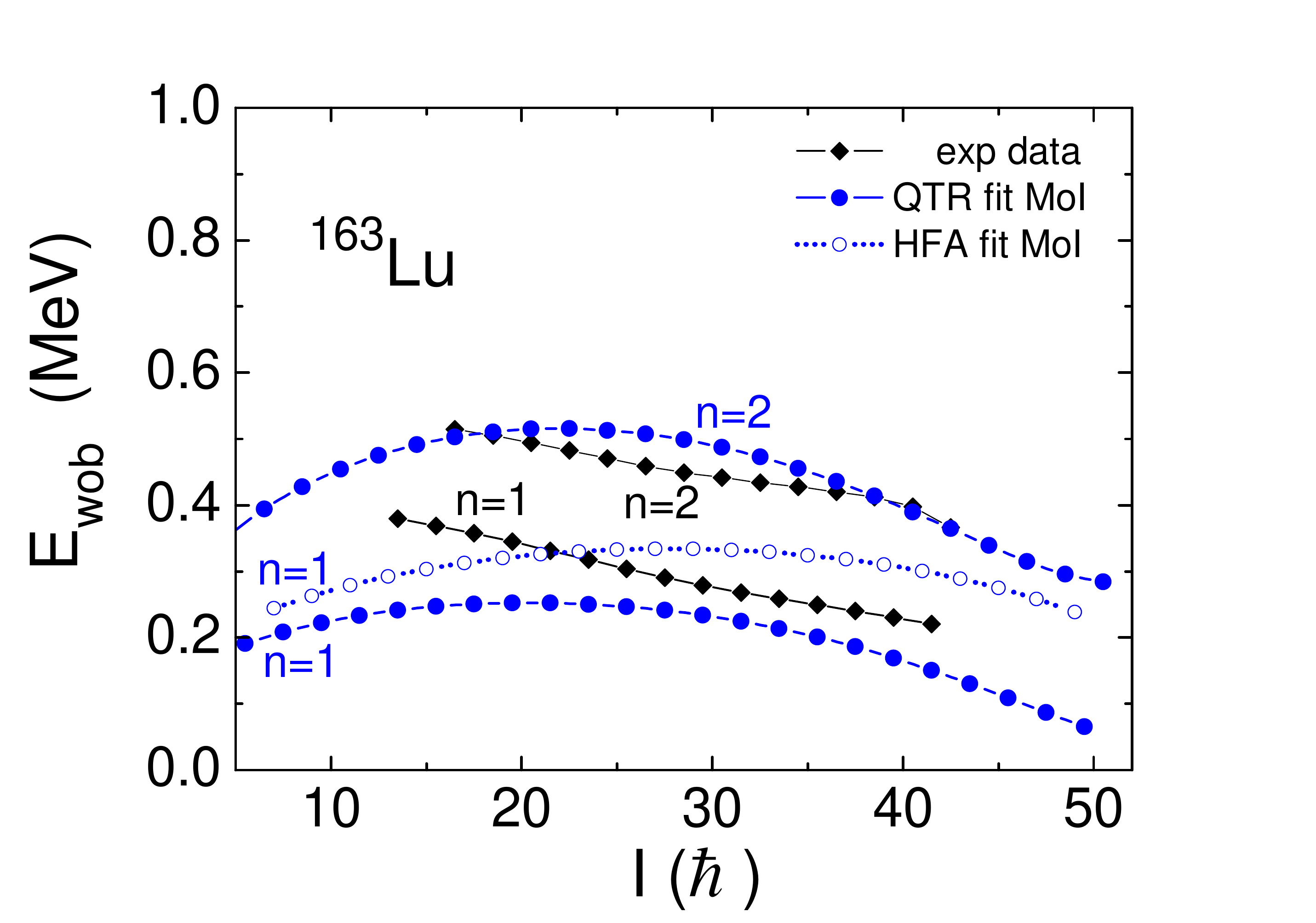} 
\hspace{-0.4cm} 
\caption{\label{lu163TPRresult} (Color online) 
Excitation energies of the n=1 and n=2 $\pi i_{13/2}$ wobbling bands in $^{163}$Lu. Solid blue lines and full dots: QTR  calculation with fitted MoI. Dotted blue line and open dots: HFA
calculation for the n=1 band with fitted MoI.
Black lines and full diamonds: Experimental data.
}
\end{figure}  

\begin{equation}
\label{HCPM}
H=h_{sqp}+ H_{core} -\kappa\sum_\mu {q_\mu^* Q_\mu},
\end{equation} 
where $h_{sqp}$  accounts for the presence of  the spherical potential and the monopole pair field,  and $H_{core}$ describes the collective motion of the 
TR core as given by Eq. ({\ref{rham}).
 The third term realizes the quadrupole-quadrupole coupling between quasiparticles and the core. 
 For completeness, we show in Appendix A   the equivalence of  the CQM Hamiltonian (\ref{HCPM})  with the  more familiar  QTR Hamiltonian (\ref{prham}).

\begin{figure}
\includegraphics[clip,height=6.5cm]{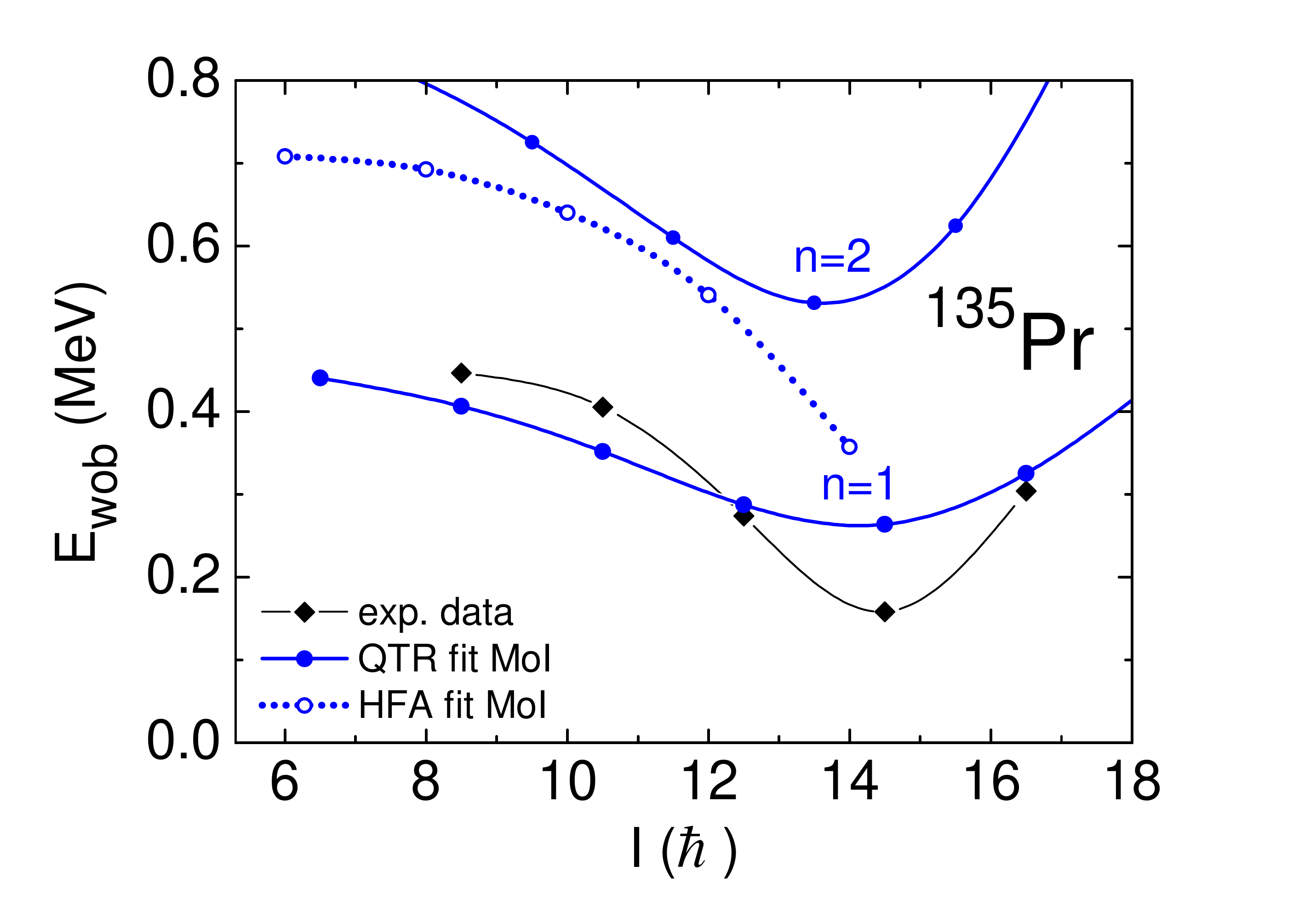} 
\hspace{-0.4cm} 
\caption{\label{pr135TPRresult} (Color online)
Excitation energies of the n=1 and n=2 $\pi h_{11/2}$ wobbling bands in $^{135}$Pr. Solid blue lines and full dots: QTR calculation with fitted MoI. Dotted blue lines and open dots: HFA calculation with fitted MoI.  
Black line and full diamonds: Experimental data.
}
\end{figure} 
The Hamiltonian (\ref{HCPM}) is
diagonalized within the KKDF frame work, which  combines the equations of motion of Kerman and Klein \cite{KK} with the projection technique
of D\"onau and Frauendorf \cite{CPM}. 
The CQM proved to be a  flexible method for coupling any model for the collective quadrupole mode of the even-even core with one quasiparticle to
describe the spectral properties of 
the considered odd-A nucleus. In our case 
the cores of $^{135}$Pr and $^{163}$Lu  are assumed to be triaxial rotors, and
the results of  our CQM are  equivalent with the ones of a QTR calculation in a deformed basis,  as e.g. the calculations in 
Ref.\,\cite{Je02,Ha02,Ha03}.  
In a first step, the simple TR problem is numerically solved, which provides the
matrices $H_{core}$ and $Q_\mu$. Since we consider the coupling to pure h$_{11/2}$ and i$_{13/2}$ quasiprotons, respectively, the term $h_{sqp}$ 
contains only the gap parameter $\Delta$ and the difference $\varepsilon-\lambda$ between the spherical single particle level $\varepsilon$ and the chemical potential $\lambda$. In the considered  
nuclides we assume the odd quasiparticles to be totally particle-like
by taking $\varepsilon-\lambda =$ 6 and 10 MeV in $^{135}$Pr and $^{163}$Lu, respectively. The chosen value $\Delta=1$ MeV of the gap is then unessential for the results of our calculations.

The input parameters needed for specifying the properties of the rigid triaxial rotor core are summarized in Tab.\ref{tab:parameter}.
The deformation parameters $\varepsilon$ and $\gamma$ are taken from the energy minima of the total routhian surfaces calculated 
with the micro-macro method.  
Actually, for  $^{163}$Lu we extracted the deformations from Fig.\,1 in Ref.\,\cite{Je02}.
The deformations of  $^{135}$Pr are identified with the minimum values in the total routhian surface shown in Fig.\,\ref{TRS135Pr},
 which we calculated with the tilted axis cranking (tac) code \cite{TAC}.
As discussed  in Ref.
\cite{Sh09},  the deformation parameters as well as the MoI obtained from cranking mean field calculations moderately depend on spin.
 This dependence is  neglected, and  the 
parameter in Tab.\ref{tab:parameter} are to be considered as average values. We are  studying  the consequences  of the spin dependence of the core parameters
in the frame work of the QCM and will report the results in a forthcoming paper.
We investigated three parameter sets for the MoI. The first set was obtained by freely adjusting of the MoI to achieve 
optimal agreement with the experimental energies of the wobbling bands.  For the second set we adopted to ratios of the MoI 
obtained by means of tac model \cite{TAC}.
For the third set we used   the calculated triaxiality 
  $\gamma$ in order to determine the ratios between the MoI according to hydrodynamic 
model  ${\cal J}_{k=1,2,3} = {\cal J}_o \sin^2(\gamma -\,2\pi k/3)$. For both the second and third set
the  scaling factor ${\cal J}_o$ for the MoI was determined recursively by adapting ${\cal J}_o$  to the energies of the n=0
 zero-phonon band of the wobbler. 
The results of the QTR calculations are presented in  Figs. \ref{lu163TPRresult} - \ref{lu163bm1fit} and Tab.\,\ref{mixing ratios}.
They are compared with the available experimental data and in some cases with the HFA.

\subsection{Quality of the HFA}\label{sec:qualHFA}
\begin{figure}
\includegraphics[clip,height=6.5cm]{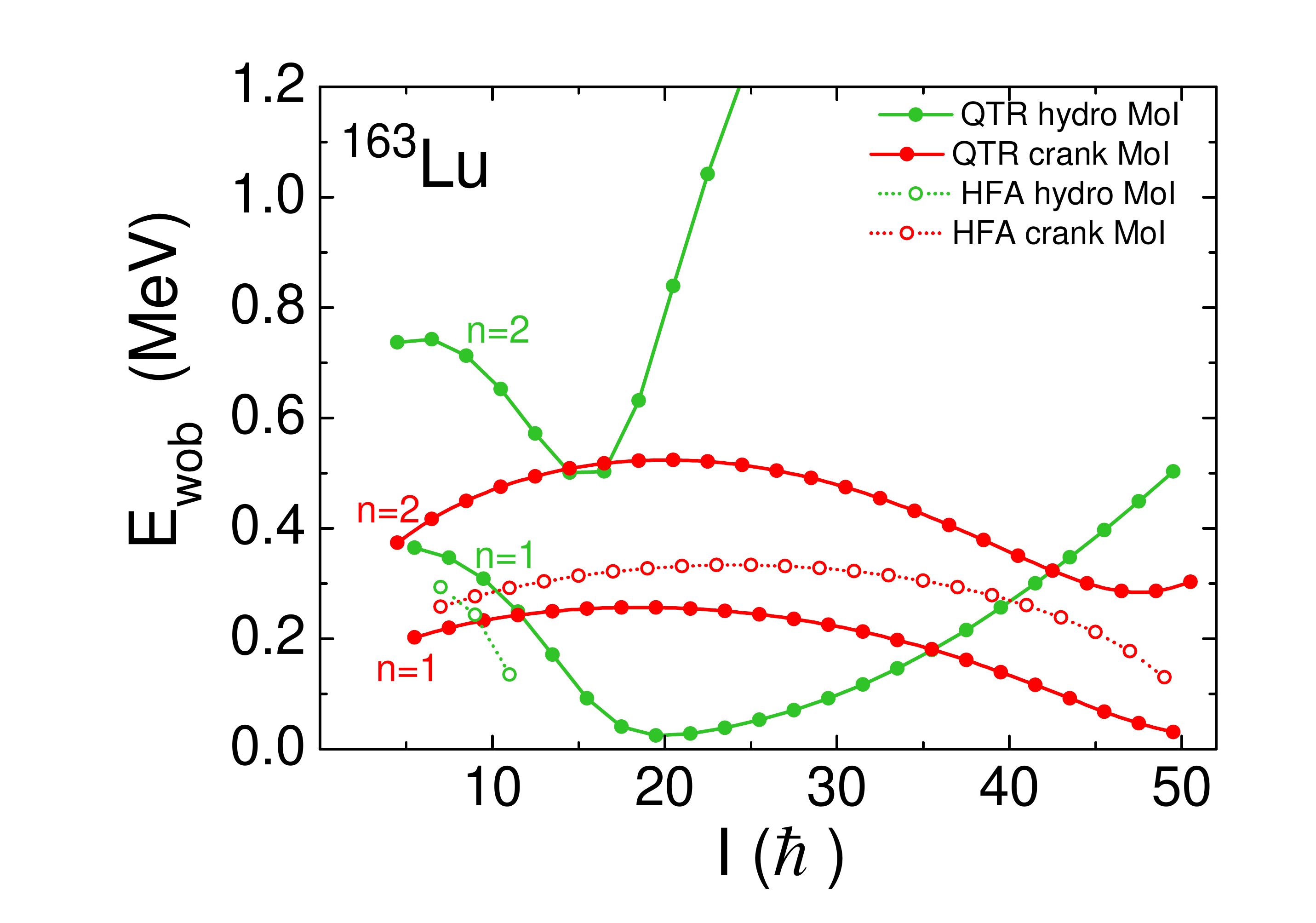} 
\hspace{-0.4cm} 
\caption{\label{lu163ewobcrank} (Color online) Excitation energies of the n=1 and n=2 $\pi i_{13/2}$ wobbling bands in $^{163}$Lu. Solid red lines and full dots: QTR with cranking MoI, solid green lines and full dots: QTR with hydrodynamic MoI. Dotted lines and open dots: HFA with
cranking and hydrodynamic MoI, respectively.
}
\end{figure}

Figs. \ref{lu163TPRresult}, \ref{lu163be2fit} and \ref{lu163bm1fit} compare QTR energies and transition probabilities for $^{163}$Lu with the HFA approximation. The HFA 
reproduces the full QTR in a fair way.   Fig.\,\ref{pr135TPRresult}    shows that   the differences between the QTR  and  HFA are more significant for   $^{135}$Pr.
This is expected, because  the $\vec j$ of the odd proton is less firmly aligned with the s-axis in $^{135}$Pr, which has a small  deformation ($\varepsilon=0.16$), compared to 
 $^{163}$Lu, which has a   much larger  deformation 
($\varepsilon=0.4$). Although the scale of the wobbling frequency in  $^{135}$Pr is substantially overestimated,  the instability of the transverse
wobbler in HFA  lies close to the minimum of the wobbling frequency of the QTR calculation. The merging of the zero- and one- phonon bands into a $\Delta I=1$ band after
the HFA instability does not occur for the QTR calculation. Instead, the relatively weakly coupled proton realigns with the $m$-axis, and the wobbler changes from
the transverse into the longitudinal mode. Fig.\,\ref{lu163ewobcrank} also shows the QTR calculation for $^{163}$Lu using the  hydrodynamic ratios between the MoI. 
As expected  from Eq.(\ref{om2wob}), the larger ratio of ${\cal J}_s/{\cal J}_m=2.34$ (as compared to 1.14 for the cranking ratios) down shifts the instability to $I=20$.
In the spin range  $20 <I< 30$ the two signature sequences are very close in energy. The larger deformation of $^{163}$Lu delays  the realignment of the odd
proton to higher a.m., whereas in  the less  deformed nuclide $^{135}$Pr it  occurs already at the instability. For $n=1$, 
the HFA for the ratio $B(E2,I$$\rightarrow$$I$$-$$1$$)/B(E2,I$$\rightarrow$$I$$-$$2$)
 somewhat overestimates the scale of the QTR values but nicely  follows the down trend with $I$. For $n=2$ the HFA estimate (two times the $n=1$ value)
 largely overestimates the QTR values  (see Fig. \ref{lu163be2fit}).

\subsection{Comparison with experiment}

The decreasing experimental wobbling frequency    classifies both  $^{163}$Lu and $^{135}$Pr  as transverse wobblers.  $^{163}$Lu  
has a quite long band with an almost linear fall-off of the frequency. The sequence in $^{135}$Pr is short, and the wobbling frequency turns up at its end. 
As already discussed, the difference is the consequence of  the different  deformation $\varepsilon$.
The considered bands in $^{163}$Lu are based on a 
highly deformed shape with $\varepsilon=0.4$. The  bands in $^{135}$Pr belong to
a weakly deformed shape with $\varepsilon=0.16$. This results in a factor of three between the  size of the  moments of inertia  
 and consequently in the scale of the a.m. Accordingly, one expects the strongly deformed $^{163}$Lu to be a 
better case than the less deformed  $^{135}$Pr for describing the wobbling motion in terms of a rigid triaxial rotor. 

The general trend of the decreasing wobbling frequency
is reproduced in our calculations with the QTR and the HFA model.
According to the HFA approximation, the wobbling band terminates at the critical spin value where the a.m. 
of the triaxial rotor changes from principal axis to tilted axis rotation. One notices that this change of the a.m. coupling is born out by the QTR calculations. 
The observed kink of the band in $^{135}$Pr at $I=29/2 $  can be related to the transition. In $^{163}$Lu the predicted end of transverse wobbling 
regime at $I=103/2$ has not been reached at $I=83/2$ in experiment. 

 Figs.\,\ref{lu163TPRresult}, \ref{lu163ewobcrank} -
\ref{lu163be2fit} compare  the excitation energies of the wobbling bands in $^{163}$Lu and $^{135}$Pr obtained from 
QTR model with cranking and hydrodynamic MoI  with the ones obtained with the MoI 
adjusted to best agreement with experiment (cf. Tab.\ref{tab:parameter}). 
 In the case of  $^{163}$Lu, the microscopically calculated cranking MoI  are very close to the fitted values and give a  satisfactory agreement with the 
experimental data. The cranking model works well in the high spin region of  well deformed nuclei. In contrast,  the hydrodynamic MoI  lead to a
much too early instability at $I=20$, where the wobbling frequency  starts increasing again. As discussed in Sec.\,\ref{sec:qualHFA},
this is the consequence of the large ratio of ${\cal J}_m/{\cal J}_s$ = 2.34.
In $^{135}$Pr the calculations  with the cranking MoI and with the hydrodynamic MoI give a too small critical spin (minimum), which reflects the large ratio
${\cal J}_m/{\cal J}_s$ = 2.4 and 3.3, respectively. The ratio 1.6 between the fitted MoI places the minimum at the right value of $I$. The QTR reproduces the observed 
increase of the wobbling frequency above $I=14$. It is caused by the reorientation of the odd proton from being aligned with the short axis toward the medium axis, which 
corresponds to a transition from transverse to longitudinal wobbling (c.f. discussion in Sec.\,\ref{sec:qualHFA}). 
Experimentally, the minimum is more pronounced than in the QTR calculations. One reason is that the experimental function $I(\omega)$ of the yrast 
band shows a back bend at $I=14$, which is not accounted for by the TR core of the QTR.

The two-phonon ($n=2$) wobbling band  has been identified in $^{163}$Lu, which the QTR calculation places close at the 
observed excitation energy. The light convex bending of both the one- and two-phonon
 bands obtained by the QTR calculation is not seen in the experimental sequences of  $^{163}$Lu. 
 No candidate for a  
\begin{figure}
\includegraphics[clip,height=6.5cm]{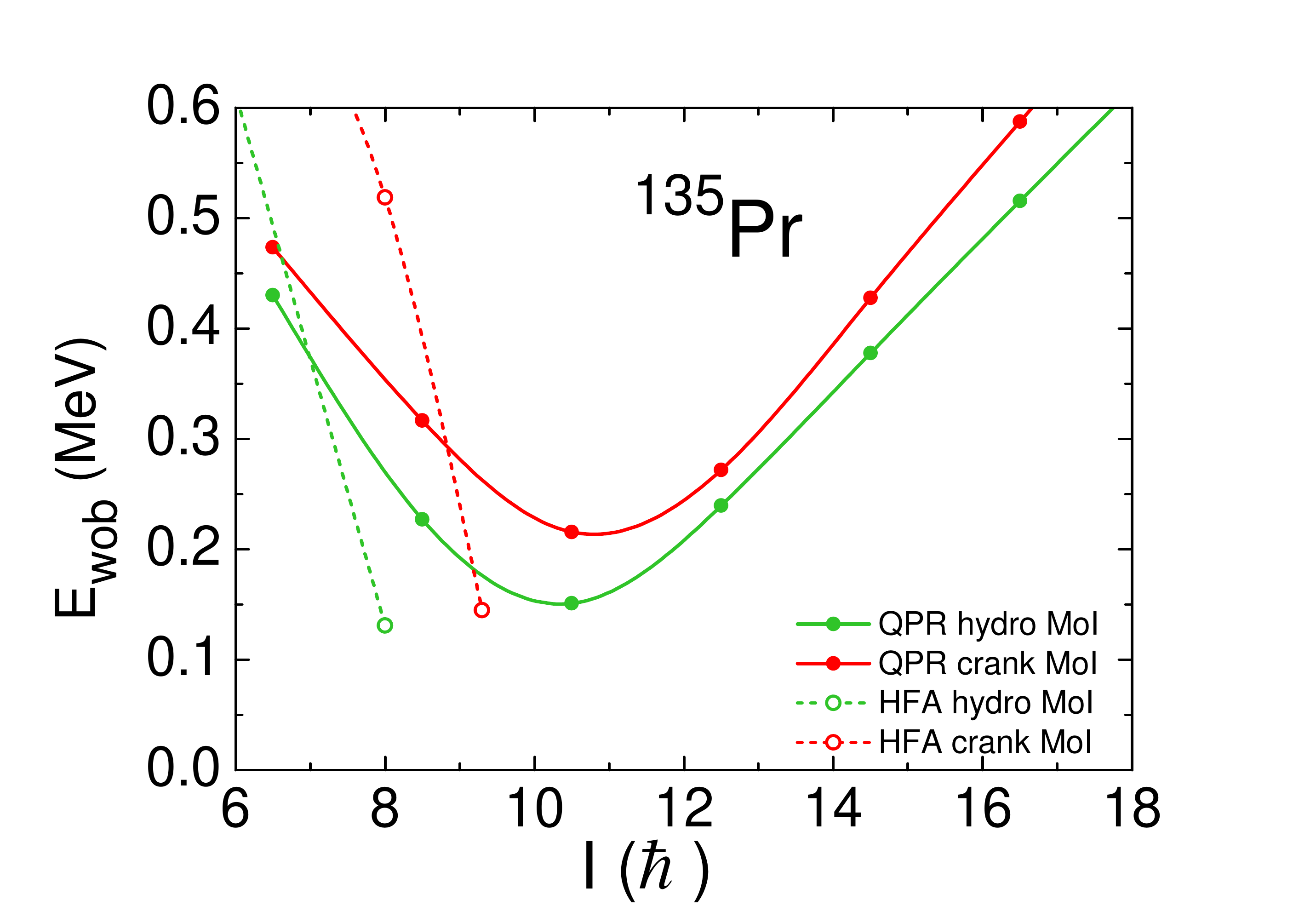} 
\hspace{-0.4cm} 
\caption{\label{pr135ewobcrank} (Color online) Excitation energies of the $\pi h_{11/2}$ wobbling band in $^{135}$Pr. Solid red lines and full dots: QTR with cranking MoI, solid green lines and full dots: QTR with hydrodynamic MoI. Dashed lines and open dots: HFA with
cranking and hydrodynamic MoI, respectively.
}
\end{figure}  
\begin{figure}
\includegraphics[clip,height=6.5cm]{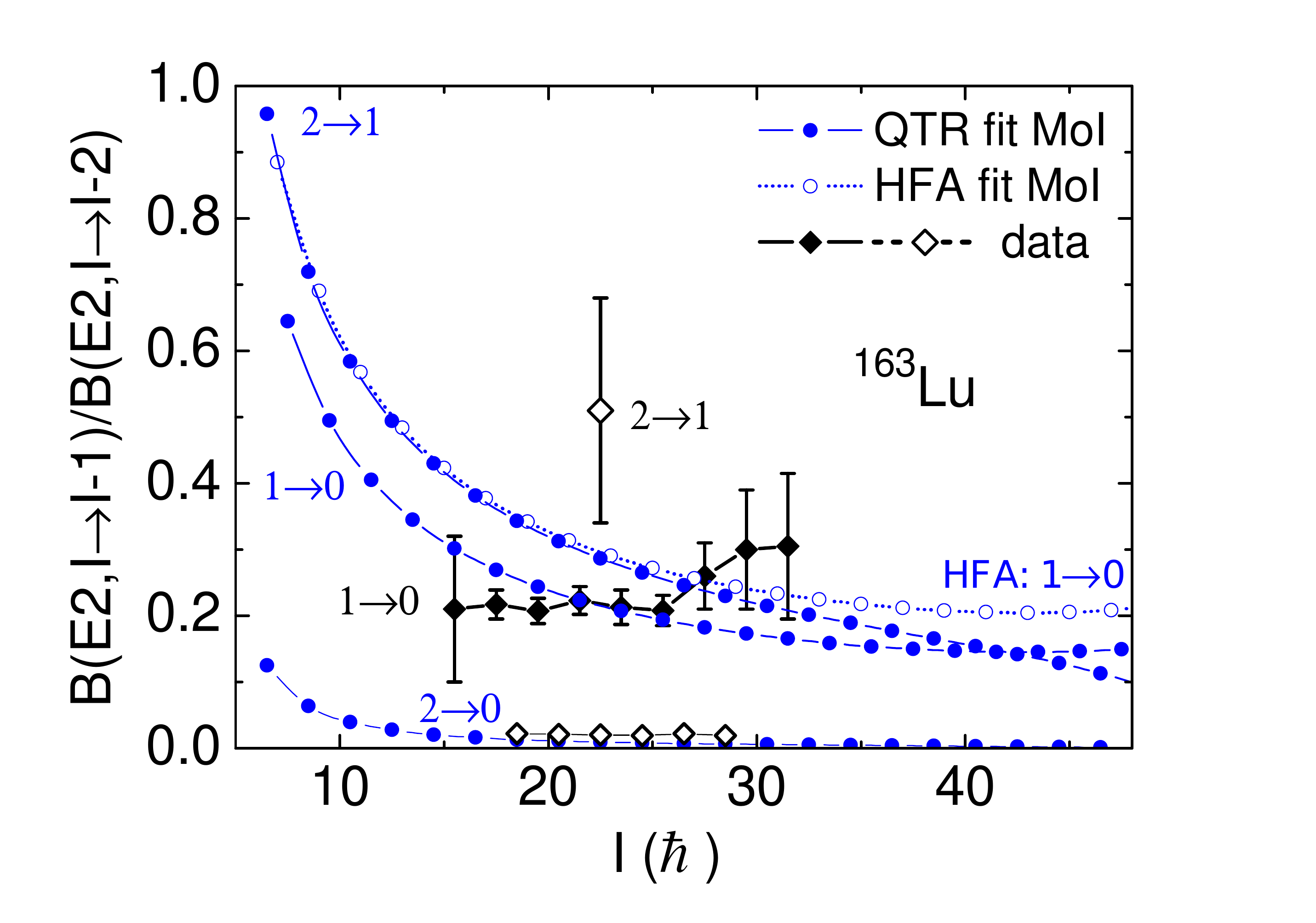} 
\hspace{-0.4cm} 
\caption{\label{lu163be2fit} (Color online) B(E2) ratios of the connecting 
to in-band transitions $n=1,2\rightarrow n=1,0$ of the wobbling bands in $^{163}$Lu. Solid blue line and full dots: QTR with fitted moments of inertia, dashed blue line and open dots: HFA with fitted moments of inertia. Black: Experimental data. The numbers indicate the actual transition $n\rightarrow n'$.
}
\end{figure}  
\begin{figure}
\includegraphics[clip,height=6.5cm]{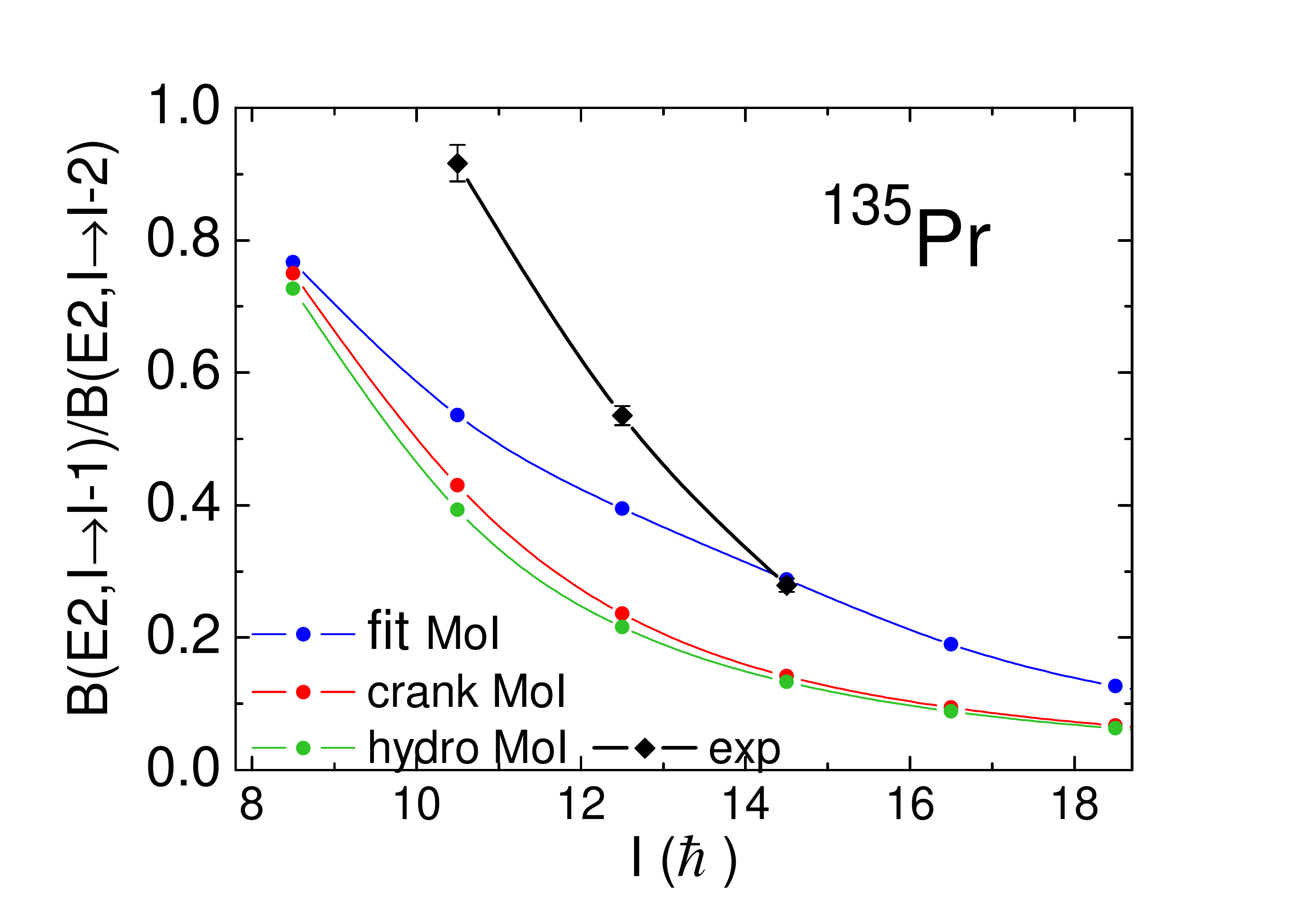} 
\hspace{-0.4cm} 
\caption{\label{pr135be2fit} (Color online) B(E2) ratios of the connecting to in-band  transitions $n=1\rightarrow n=0$ of the wobbling band in $^{135}$Pr. Solid blue line: QTR calculated with fitted moments of inertia, red (green) line:  with cranking  (hydrodynamic) moments of inertia. 
Black: Experimental  data (cf.\,Tab.\ref{mixing ratios}).
}
\end{figure}  
\begin{figure}
\includegraphics[clip,height=6.5cm]{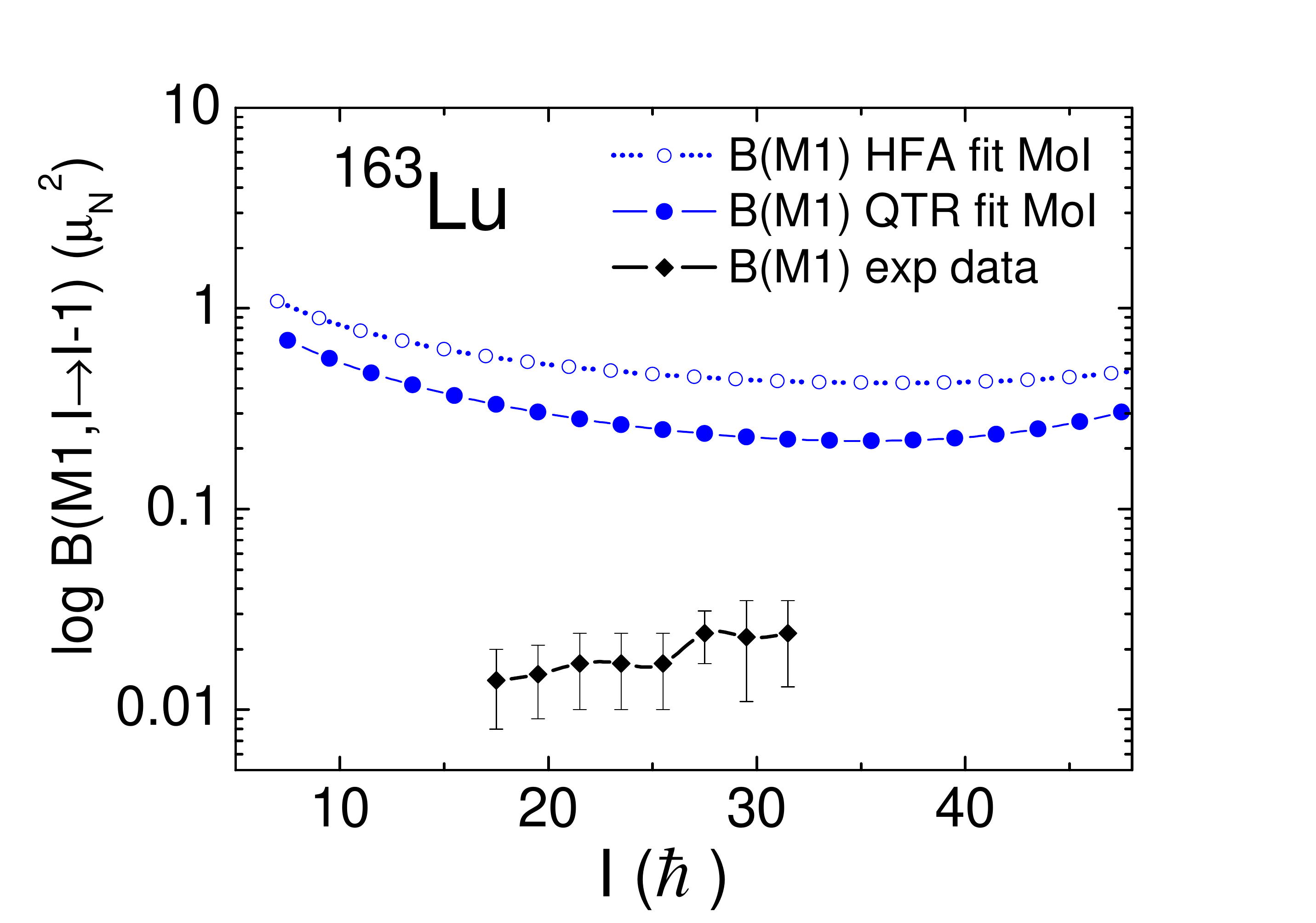} 
\hspace{-0.4cm} 
\caption{\label{lu163bm1fit} (Color online) B(M1,$I$$\to$$\,I$-1) of the connecting  transitions $n=1\rightarrow n=0$ of the wobbling band in $^{163}$Lu. Solid blue line and full dots: QTR with fitted moments of inertia, dashed blue line and open dots: HFA with fitted moments of inertia. Black: Experimental  data.
}
\end{figure}  
\begin{table}
\caption{ \label{mixing ratios} B-values and  ratios  r$_{M1}$ =
B(M1)$_{con}$/B(E2)$_{in}$ and r$_{E2}$ = B(E2)$_{con}$/B(E2)$_{in}$
of the wobbling states in $^{135}$Pr. The suffices $in$ and $con$ refer to the
in-band $\Delta$$I$=$2$ and  $\Delta$$I$=$1$ transitions that connect the bands, respectively. 
The B-values and the calculated ratios r$^{cal}$ result from our QTR 
calculations with the fitted MoI.  The tentative 
experimental ratios r$^{exp}$ were made available prior publication by J. Matta {\em et al.} \cite{Garg}.}
  \begin{ruledtabular}
\begin{tabular}{cccc}
spin  $I$   & B(E2)$_{in}^{cal}$  &B(E2)$_{con}^{cal}$  &B(M1)$_{con}^{cal}$ \\
\hline
17/2 & 0.637&0.488 & 0.243\\
21/2 &0.744&0.399&0.187 \\
25/2 &0.838&0.331&0.167    \\
29/2&0.930&0.267&0.167 \\
33/2&1.023&0.194&0.155\\
\end{tabular}
\begin{tabular}{ccccc}
spin  $I$ & r$_{M1}^{cal}$& r$_{M1}^{exp}$& r$_{E2}^{cal}$& r$_{E2}^{exp}$\\
\hline
17/2 & 0.381& - & 0.767 & - \\
21/2 &0.252 & 0.130$\pm$ 0.011 &  0.536 & 0.92$\pm$ 0.03  \\
25/2 &0.199 &0.021$\pm$ 0.004&0.395  &0.535$\pm$ 0.01  \\
29/2& 0.180& 0.010$\pm$ 0.002 & 0.288& 0.28$\pm$ 0.01  \\
33/2& 0.151&- & 0.190& -  \\
\end{tabular}
\end{ruledtabular}
\end{table}
two-phonon has been found in $^{135}$Pr.

As  can be seen in Fig.\,\ref{lu163TPRresult}, the QTR energy of the two-phonon 
is about  twice the energy of the one-phonon bands for $I< 30$, 
whereas this ratio is considerably lower in the experimental band.
So far, we have no explanation for this discrepancy. For larger values of $I$, the QTR ratio increases.
This is explained as follows. The one- phonon band has opposite signature than the zero-phonon band.
Because of their different symmetry the two bands may approach and even cross. The two-phonon
band has the same signature as the zero-phonon band, and the two bands can mix and repel each other.
The onset of the repulsion is seen around $I=50$. The experiment also shows an increasing of the ratio of the two- and one-phonon energies. 

  The ratios $B(E2,I$$\rightarrow$$I$$-$$1$$)/B(E2,I$$\rightarrow$$I$$-$$2$) 
for the transitions connecting the bands (con)   and the in-band transitions (in) in $^{163}$Lu are shown in Fig.\,\ref{lu163be2fit}. 
 The QTR calculations reproduce the strong connecting E2- transitions observed in experiment,
which are  the evidence for the collective nature of wobbling excitations.   The  calculated ratios 0.3-0.2 are in accordance with  experiment  \cite{Je01}.
However, the experimental ratio is weakly increasing within the observed spin interval $I=31/2-63/2$, whereas  the QTR calculation gives a  slightly decreasing ratio.
 The HFA shows the same tendency as QTR.
The measured ratio 
$B(E2,I$$\rightarrow$$I$$-$$1$$)/B(E2,I$$\rightarrow$$I$$-$$2$)=  0.5$\pm$0.15
at $I=45/2$ is about twice as large as the ratio of the corresponding 
n=1 to n=0 transition which supports the two-phonon nature of the
upper band. In the QTR calculation the corresponding ratio between n=2 and 
n=1 out-of-band transitions is with about 1.3 too low for a clear two-phonon
structure of the calculated n=2 band.   

  The ratios $B(E2,I$$\rightarrow$$I$$-$$1$$)/B(E2,I$$\rightarrow$$I$$-$$2$) 
shown in Fig.\,\ref{pr135be2fit} for $^{135}$Pr are obtained with the QTR model
using  the three parameter sets for the MoI  in Tab.\ref{tab:parameter}. 
The ratios slope down with increasing spin. 
The calculation with the fitted moments of inertia predicts 
 the strongest interband E2 transitions, corresponding to a ratio  of 0.5-0.3 within the spin interval $I$\,=10-14.  The tentative 
 experimental ratios \cite{Garg} are larger and decrease more rapidly.
 
 Fig.\ref{lu163bm1fit}  displays the $B(M1)$ values of the connecting transitions 
$I$$\to$$\,I$$-1$ of the one-phonon wobbling band in $^{163}$Lu. The measurements
\cite{Je01} found very  small M1 admixtures to the non-stretched E2 transitions. The QTR and the HFA calculation also predict only a weak M1 admixture. 
However,  the relatively flat curve with a minimum of 
about 0.2\,$\mu_N^2$  overestimates the measured $B(M1)$ values by more than a factor of 10. 
The QTR calculations with the cranking and hydrodynamic moments of inertia gave similar  $B(M1)$ 
values. 
The QTR calculations for the wobbling band in $^{135}$Pr predict  $B(M1,I\rightarrow I-1) \approx$\,0.2\,$\mu_N^2$ for each of the three sets of MoI.
The recent  experiment  by J. Matta {\em et al.}  \cite{Garg} provided the tentative ratios  r$_{M1}$ = B(M1)$_{out}$/B(E2)$_{in}$  in Tab.\ref{mixing ratios}, which are a factor
of 10 smaller than the  calculated ones. These ratios suggest that the M1 strength obtained with QTR is too large
for $^{135}$Pr as well. 

The reproduction of the very small $B(M1)$ values is a problem for the QTR description of transverse wobbling. 
The mechanism for generating the M1 radiation is simple in the HFA.
  The magnetic moment $\vec \mu=g_j\vec j$  of the high-j particle is aligned with the s-axis. It 
wobbles together with the rotor  generating the M1 radiation. The transition rate is given by the squared amplitude of this 
oscillation, which is determined by the  wobbling amplitude of the rotor and 
the length of $\vec \mu$.    As seen in Fig. \ref{lu163bm1fit}, the HFA gives somewhat larger $B(M1)$ than QTR, 
which can be attributed to the HFA assumption of rigid alignment of the high-j particle with the s-axis. In the case of QTR,
 the coupling of the  high-j particle is not rigid. Its $\vec \mu$  does not completely follow the motion of the rotor,
which reduces the amplitude of its own wobbling motion and, as a consequence, the intensity of the M1 radiation. 
 Hence, the $B(M1)$ values of QTR reflect the degree of alignment of the high-j particle with the short axis, 
 i. e. the transverse character of the wobbling motion. We assume that the experimental $B(M1)$ values are substantially smaller 
 than the calculated ones, because there are additional  couplings between the rotor core and the quasiparticle that the QTR
 does not take into account (The CQP model considers only the coupling to the deformed quadrupole field.)
 This conjecture is supported by our study of transverse wobbling in the framework of QRPA in a forthcoming paper \cite{QRPA}

\section{Summary}

Studying the classical orbits of the angular momentum vector of a triaxial rotor we have demonstrated that
  the presence of a high-j  quasiparticle, which rigidly aligns its angular momentum $\vec j$  with
one of the principal axes, drastically changes the motion of the coupled system.
 Two types of wobbling motion appear: the longitudinal and the transverse, depending, respectively, on 
 whether the quasiparticle $\vec j$ is {\em aligned with} the  axis of 
the largest MoI ({\em longitudinal wobbler}) or is oriented {\em perpendicular}  to this axis ({\em transverse wobbler}). 
The assumption that the quasiparticle $\vec j$ is rigidly aligned with one of the principal axes (Frozen Alignment - FA) allowed us to derive simple analytical expressions 
for the wobbling frequency and E2 and M1 transition rates 
in analogy to the well known formulae obtained by applying the harmonic approximation  to the motion of 
the triaxial rotor \cite{Bo75} (Harmonic FA -HFA). Our simple  HFA expressions help to understand why for the longitudinal alignment the wobbling frequency 
monotonically increases with the total angular momentum
  whereas it decreases  for transverse alignment. There is a  critical angular momentum where the transverse wobbling regime ends and the one- and zero- phonon
  bands merge into one $\Delta I=1$ sequence. The simple HFA  expressions provide a classification scheme for the wobbling motion and a
  qualitative understanding of the results obtained in the framework of the more realistic Quasiparticle+Triaxial Rotor (QTR) model. 
  All strongly deformed wobbling bands observed at high spin in the Lu and Tm isotopes carry the signature of transverse wobbling.
   
 We studied the excitation energies and the electromagnetic E2 and M1 transition rates of transverse wobbling states  in $^{163}$Lu and $^{135}$Pr in the 
 framework of the QTR. The deformation parameters of the rotor were calculated by means of the micro-macro method.
 The three moments of inertia of the rotor were considered as free parameters, which were adjusted to fit the experimental
 energies of the zero- and one-phonon wobbling bands.  
 Good agreement with the measured energies and E2 strengths was found for the high spin wobbling bands in $^{163}$Lu, 
 which has a strongly deformed  triaxial shape. The signature of transverse wobbling, the decrease of the wobbling frequency with angular momentum, 
  was reproduced. In accordance with experiment, the predicted critical spin of $I\approx$\,50  of the transverse wobbling band is higher than 
  the  observed spins. The ratios between the three moments of inertia determined by the fit turned out to be close the ones calculated
 by means the cranking model. Assuming  the ratios for irrational flow resulted in a much too low critical spin.
Because the moments of inertia  of the weakly deformed $^{135}$Pr are  smaller by a factor of about three, the wobbling bands appear  lower spin. 
Fair agreement agreement of the QTR results  with the measured energies and E2 strengths was found as well. At low spin the wobbling mode is transverse.
At the critical spin of  $I=29/2$ the wobbling mode changes from transverse to longitudinal, which is caused by a realignment of the 
of the h$_{11/2}$ proton from the short to the medium axis. The ratios between the fitted moments of inertia did not agree with the ones obtained by assuming 
irrotational flow nor with the ones calculated by means the 3D-cranking model, both of which gave a too low critical spin. However for all three ways of determining 
the moments of inertia, the medium axis has the largest one, followed by the short axis, and the long axis.

In summary, the concept of transverse wobbling provides a natural explanation for the decrease of the wobbling frequency with increasing
angular momentum and the enhanced E2 transitions between the wobbling bands. It is decisive that the ratios between the three moments of inertia
of the triaxial rotor and the orientation of the odd quasiparticle  are in qualitative agreement with microscopic calculations based on the cranking model.

\appendix
\section{}
The relation between the QTR Hamiltonians in the laboratory and intrinsic frames of reference is established
by transforming the coupling term $\sum_\mu {q_\mu^* Q_\mu}$ to the
principal axis (PA) system of the core. Because this term is
rotational invariant we obtain  
\begin{equation}\label{qqterm}
\sum_\mu {q_\mu^* Q_\mu} =\bar q_0 \bar Q_0 + (\bar q_2+\bar q_{-2})\bar Q_2.
\end{equation} 
where $\bar Q_0$ and $\bar Q_2$ are the non-zero core quadrupole moments of the triaxial core. 
They can up to a scaling factor $f$ be parametrized in terms of $\varepsilon$ and $\gamma$ as $\bar Q_0=f\,\varepsilon\cos{\gamma}$ and $\bar Q_2=f\,\varepsilon\cos{\gamma}/\sqrt{2}$.
Defining the particle quadrupole moments as $q_\mu=r^2\,Y^2_\mu$
 the coupling term takes the familiar form of a triaxially deformed potential 
\begin{equation}\label{potential}
\sum_\mu {q_\mu^* Q_\mu}= f \, r^2(\varepsilon\cos{\gamma}\bar Y^2_0 +
\varepsilon\cos{\gamma}/\sqrt{2} ( \bar Y^2_2 + \bar Y^2_{-2} )).
\end{equation} 
The appropriate scaling factor $f$ for obtaining 
the standard form of the deformed potential   
in the resulting Hamiltonian (\ref{prham}) is given by 
\begin{equation}
\kappa\,f= \frac{2}{3}\sqrt{\frac{4\pi}{5}}\,\,\hbar\omega_\circ
\approx 1.057\,\hbar\omega_\circ
\end{equation} 
where $\hbar\omega_\circ = 41 A^{-1/3} $ MeV is the oscillator energy constant.
 Inserting the resulting coupling term in Eq.\,(\ref{HCPM}) we obtain the deformed quasiparticle term $h_{dqp}$   of the QTR Hamiltonian (\ref{prham}).
Expressing in  the core part $H_c$ the core angular momentum $\vec R=\vec J-\vec j$ in terms of the total a.m. and the particle a.m. gives the 
QTR Hamiltonian (\ref{prham}). The explicit transformation of the wave functions between the laboratory and intrinsic frames can be found in   \cite{Bo75}.
The coupling strength $\kappa$ is related to the 
deformations $\varepsilon$ and $\gamma$ of the quasiparticle-core system by 
\begin{equation}
\label{kappa}
\kappa <0||Q||2> =\hbar\omega_\circ\varepsilon\cos\gamma , 
\end{equation} 
where $<0||Q||2>$ means the reduced matrix element of the core 
quadrupole operator taken between the 0$^+$ ground state and the first 2$^+$ state of the core.

\begin{acknowledgements}
  We thank J. Matta and U. Garg for providing us the tentative results of their measurements. 
  The work was  upported by the DoE Grant DE-FG02-95ER4093.
  \end{acknowledgements}

\end{document}